# DNA origami catenanes templated by gold nanoparticles

*Andreas Peil, Pengfei Zhan, and Na Liu**

Andreas Peil, Dr. Pengfei Zhan, and Prof. Dr. Na Liu
Max-Planck-Institute for Intelligent Systems
Heisenbergstr. 3, 70569 Stuttgart, Germany

Andreas Peil, and Prof. Dr. Na Liu
Kirchhoff-Institute for Physics
Im Neuenheimer Feld 227, 69120 Heidelberg, Germany
E-mail: na.liu@kip.uni-heidelberg.de

Mechanically interlocked molecules have marked a breakthrough in the field of topological chemistry and boosted the vigorous development of molecular machinery. As an archetypal example of the interlocked molecules, catenanes comprise macrocycles that are threaded through one another like links in a chain. Inspired by the transition metal-templated approach of catenanes synthesis, in this work we demonstrate the hierarchical assembly of DNA origami catenanes templated by gold nanoparticles. DNA origami catenanes, which contain two, three or four interlocked rings are successfully created. In particular, the origami rings within the individual catenanes can be set free with respect to one another by releasing the interconnecting gold nanoparticles. Our work will set the basis for rich progress towards DNA-based molecular architectures with unique structural programmability and well-defined topology.

## 1. Introduction

Mechanically interlocked molecular architectures are molecules, which are connected as a consequence of their chemical topology. Over the last decades, a variety of interlocked molecules have been created, ranging from Borromean rings[1-3], knots[4,5], catenanes[6], rotaxanes[7], pretzelanes[8,9] to Solomon-links[10] and daisy-chains[11,12]. A catenane is a molecule with two or more topologically linked macrocycles[13,14], whereas a rotaxane consists of a dumbbell-shaped axis threaded through at least one macrocycle[15]. The first catenane synthesis was realized by Edel Wasserman in 1960 at the Bell Laboratories[13,14]. Later in 1967, Harrison *et al.* reported the first synthesis of rotaxanes.[15]

For quite some time, low efficiencies and complicated procedures had been a bottleneck for the synthesis of interlocked molecules. The situation was changed after Jean-Pierre Sauvage introduced the transition metal-templated approach in 1983. This approach utilizes a transition metal cation to arrange two bidentate ligands in a tetrahedral array as a prelude to cyclization and catenane formation.[16,17] **Figure 1** shows the two representative routes. In strategy A "gathering and threading", a crescent-shaped molecule is threaded through a ring-shaped molecule via a transition metal ion and subsequently closed by another crescent-shaped molecule in a single cyclization reaction. In strategy B "entwining", two crescent-shaped molecules are linked together via a transition metal ion, followed by a double cyclization reaction to close them at once. In each case, the transition metal ion can be removed after serving its purpose, enabling the free relative movements of the two rings.

Mechanically interlocked architectures are appealing building blocks for the realization of many functional devices. Importantly, they empower broad applications in nanomaterials, nanorobotics and nanomachines.[18-21] For instance, catenanes have been exploited as switches[22], rotary motors[23,24] and sensors[25-30]. Rotaxanes have been used as molecular shuttles[31,32], switches in molecular electronics[33], control of chemical synthesis[34] and force-generating components for molecular elevators[35] and pumps[36].

An alternative pathway to create topologically complex molecular architectures is structural DNA nanotechnology, which exploits DNA as both genetic and construction material.[37-40] The first topological DNA structures were demonstrated by Seeman and coworkers, who synthesized a variety of DNA knots and Borromean rings.[41,42] It was then followed by the construction of DNA catenanes[43-46], rotary motors[47,48], a bio-hybrid nanoengine[49] and logic circuits[50].

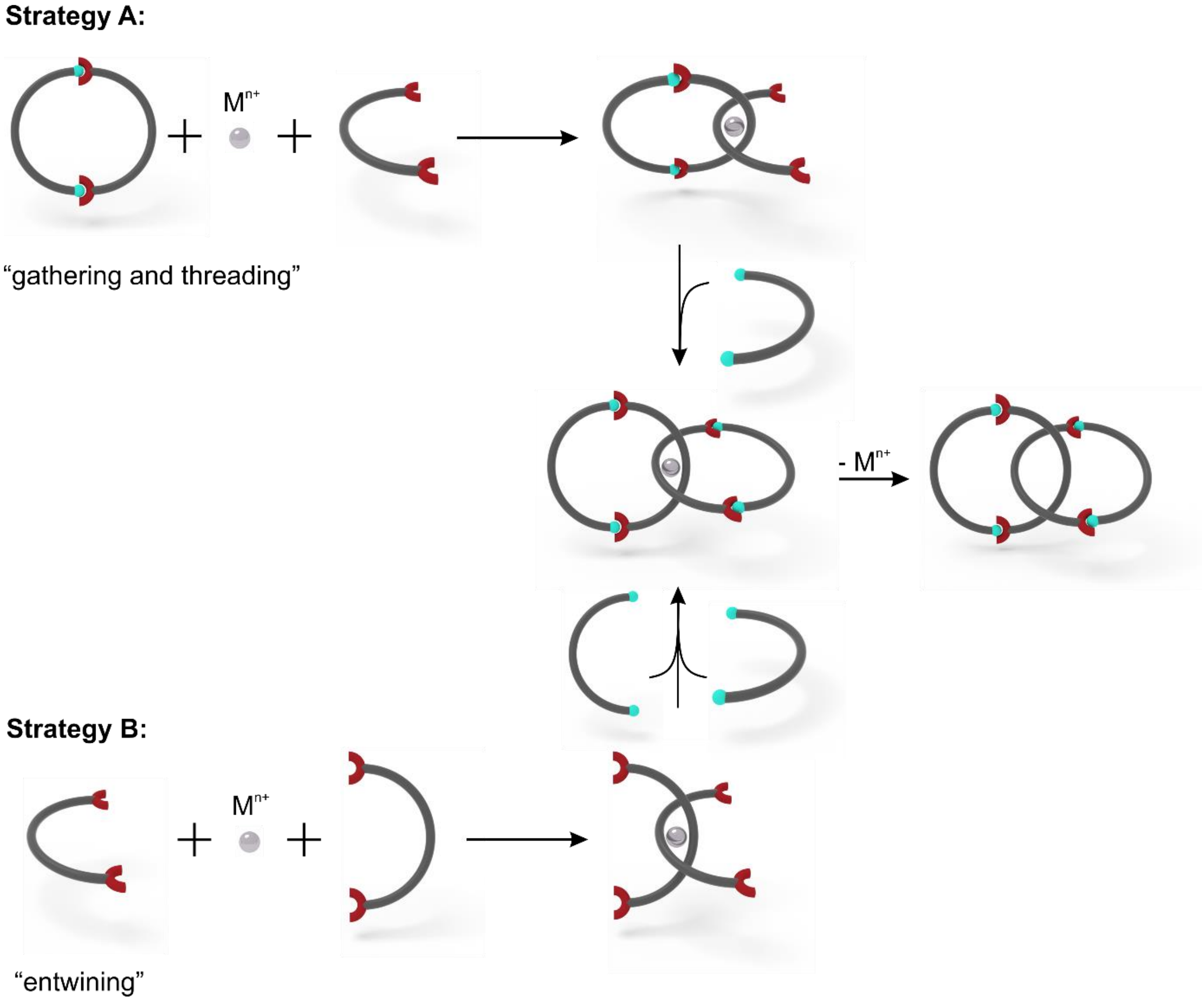

**Figure 1 | Transition metal-templated synthesis of catenanes.** In strategy A "gathering and threading", a crescent-shaped molecule and a pre-formed ring molecule are linked via a transition metal ion and subsequently closed by another crescent-shaped molecule. In strategy B "entwining", two crescent-shaped molecules are linked together via a transition metal ion, followed by a double cyclization reaction. Removal of the transition metal ion allows for the free relative movements of the two rings.

In 2010, DNA rotaxanes were reported.[51] Subsequently, chemically more rigid[52], light-switchable[53] and Daisy-chain rotaxanes[54], DNA-nanoparticle rotaxanes[55] as well as catalytically active rotaxanes[56] were demonstrated. Despite being topologically well-defined, the as-fabricated DNA catenanes and rotaxanes were in general mechanically rather flexible and floppy, because they were made of single- or double-stranded DNA. In contrast, DNA origami-based interlocked architectures possess much better structural rigidity. The first attempt to realize such catenanes was carried out by Yan *et al*., who followed an innovative scheme of molecular kirigami.[57] In 2016, Simmel *et al*. reported switchable DNA origami rotaxanes with long-range on-axis motion.[58]

In this work, we experimentally demonstrate the hierarchical assembly of DNA origami catenanes templated by gold nanoparticles (AuNPs), taking inspirations from the transition metal-templated synthesis of catenanes developed by Jean-Pierre Sauvage. DNA origami catenanes, which contain two, three or four interlocked rings are successfully created. In particular, the origami rings within the individual catenanes can be set free with respect to one another by releasing the interconnecting AuNPs through toehold-mediated strand displacement reactions.

## 2. Results and Discussion

### 2.1. Assembly of the DNA origami rings

**Figure 2** shows the schematic of the DNA origami ring formation. A long single-stranded DNA scaffold (M13) is folded by ~210 short staple strands through hybridization in a self-assembly process to form an origami monomer, *i.e*., a quarter of a ring (see Figures 2a, S1, and S2). Each monomer contains 24 curved DNA helices bundled in a honeycomb lattice with two distinct junction regions at its ends. In total, four different sets of staple strands are utilized in order to differentiate the junction regions (I, green; IIa, blue; IIb, red; III, orange) for achieving a controlled connectivity as shown in Figure 2b (see also Figure S3). An origami ring with an inner diameter of ~120 nm can be created by linking four monomers. More specifically, junction I consists of 10 head-to-head connector strands with 3-base long sticky ends. An AuNP binding site is positioned in the region of junction I (see Figure 2b). At the binding site, there are 4 capture strands of the same sequence extended from each monomer to assemble one AuNP (15 nm in diameter; for design details see Figures S4, and S5). Junction III is designed similar to Junction I, but it comprises 12 head-to-head connector strands and lacks the AuNP binding site. Junctions IIa and IIb form a pair. Junction IIa consists of 11 connector strands with 9-base long sticky ends, whereas junction IIb does not contain sticky ends (for design details, see Figures S6 and S7). Such a design scheme ensures the precise control of the origami ring connection as well as the bound AuNP number. Importantly, it also inhibits the generation of undesired side-products.

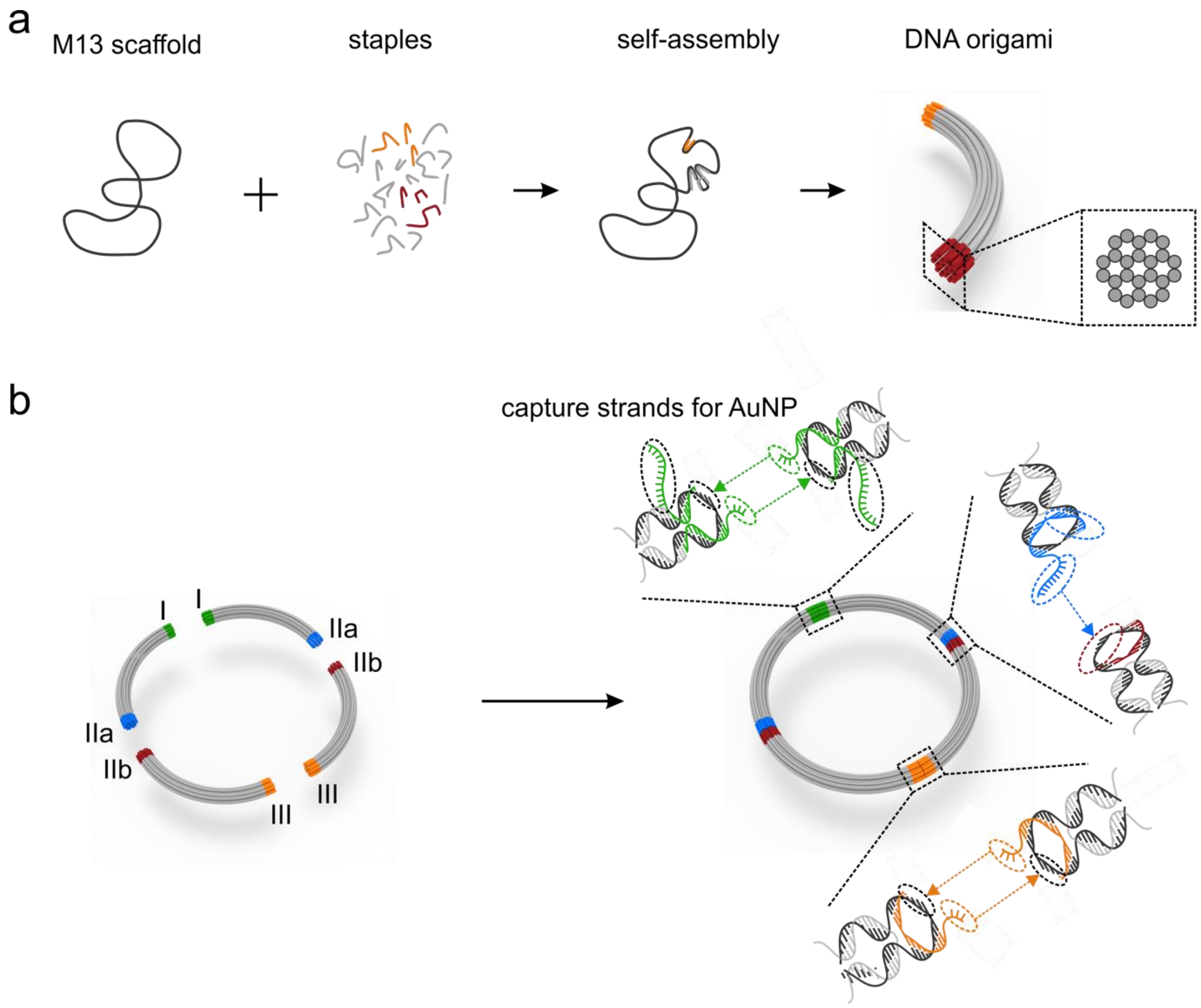

**Figure 2 | Assembly of the DNA origami rings.** (**a**) Curved DNA origami monomer is formed by folding a M13 scaffold and staple strands through a hierarchical assembly process. The monomer has two different junction regions at its ends. (**b**) Four monomers are linked together to form an origami ring with a diameter of 120 nm by DNA hybridization. Four different sets of staple strands are used to differentiate the junction regions (I, green; IIa, blue; IIb, red; III, orange) for achieving a controlled connectivity. Capture strands (green) extended from junctions I are used to anchor a AuNP.

## 2.2. Assembly of the DNA-[2]-catenanes

In resemblance to the transition metal-templated synthesis of catenanes[16], two different assembly strategies, A ("gathering and threading") and B ("entwining") are explored to create DNA-[2]-catenanes, respectively. As shown in **Figure 3a**, the assembly steps for strategy A are as follows. (i) Assembly of an origami ring by connecting four monomers (see also Figures 2b, S8, and S9). (ii) Positioning of a AuNP functionalized with complementary DNA to the origami ring through hybridization with the capture strands extended from junctions I. (iii) Formation of an origami half-ring via linking two identical monomers, which contain junctions I (AuNP-binding site) and IIa. (iv) Threading the half-ring through the origami ring by connecting to the AuNP at junctions I, giving rise to a 1.5-ring complex. (v) Cyclization of the 1.5-ring complex by another half-ring, which contains junctions IIb and III. Subsequently, a DNA-[2]-catenane is formed. Strategy B comprises fewer steps. (vi) Entwining two origami half-rings that contain junctions I and IIa around a AuNP results in the formation of an interlocked half-ring complex (see Figures S10 and S11). Subsequently, a DNA-[2]-catenane can be achieved through a double cyclization by connecting two other half-rings, which contain junctions IIb and III.

Next, strategies A and B for the assembly of the DNA-[2]-catenanes are evaluated and compared using gel electrophoresis and transmission electron microscopy (TEM) as shown in Figures 3b and 3c. The results reveal that strategy B gives rise to a higher yield than strategy A. More specifically, the gel images in Figure 3b show weak bands (see the black-dashed frames) for the 1.5-ring complexes (iv) and the DNA-[2]-catenanes (v). In contrast, the gel images exhibit clear bands for the entwined half-rings with AuNPs (vi) and the DNA-[2]-catenanes (v) as shown in Figure 3c. It is noteworthy that for each strategy the step right before the DNA-[2]-catenane formation is particularly crucial to ensure a successful assembly.

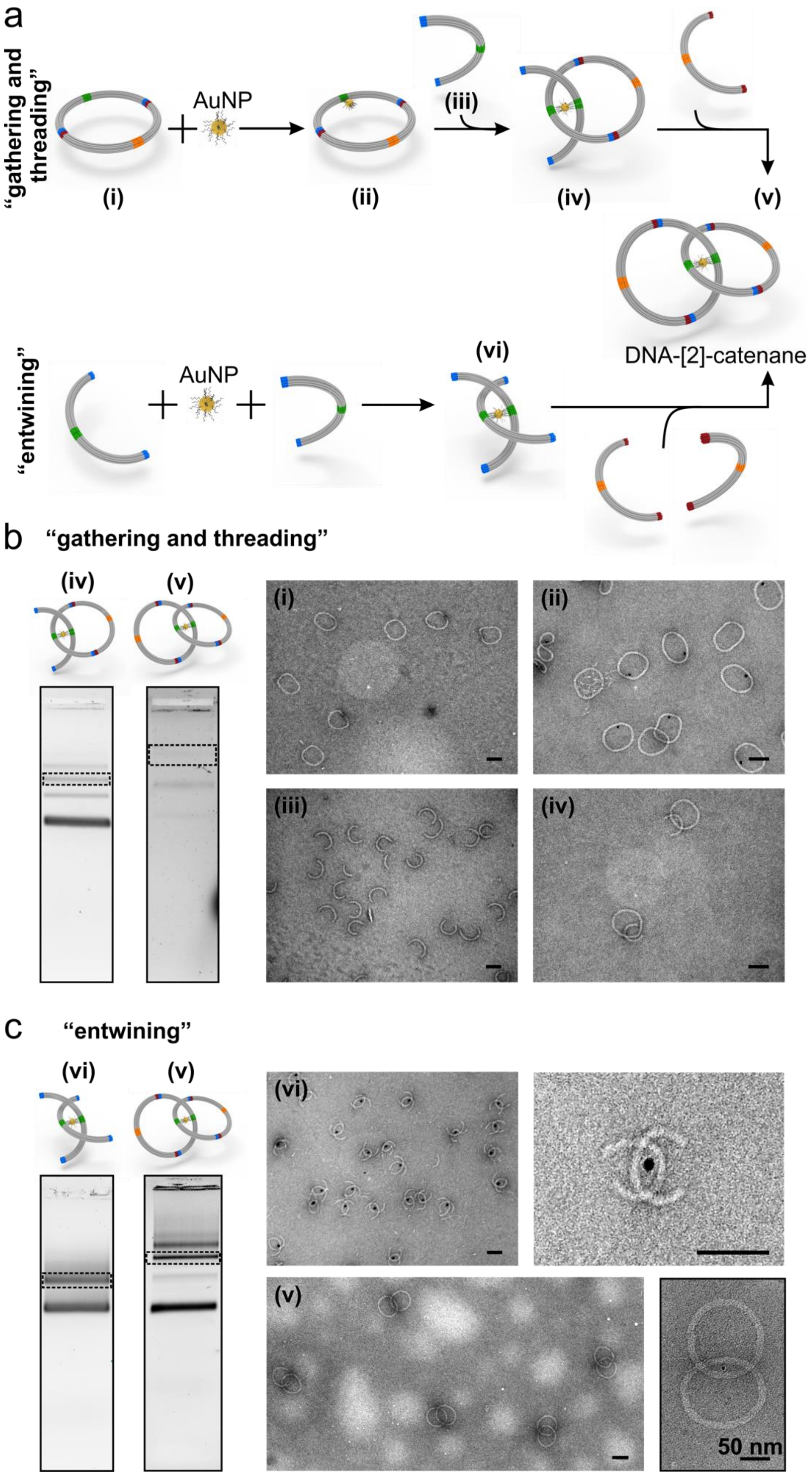

**Figure 3 | Assembly of the DNA-[2]-catenanes.** (**a**) Schematics of the "gathering and threading" and "entwining" strategies. (**b**) Agarose gel and TEM images of the structures in the "gathering and threading" strategy (**c**) Agarose gel and TEM images of the structures in the "entwining" strategy. The black-dashed frames represent the gel bands of the corresponding structures. Scale bar: 100 nm.

As presented by the TEM images in Figure 3b, the intermediate structures show relatively good yields (i, ~62%; ii, ~54%), whereas the yield of the 1.5-ring complexes (iv, ~24%) is quite low. This is likely because threading a half-ring through a complete ring is sterically hindered. As a result, creation of the DNA-[2]-catenanes by further cyclization of the 1.5-ring complex with another half-ring becomes challenging. On the other hand, the TEM image of the interlocked half-ring complexes in Figure 3c reveals a good yield (vi, ~50%), which leads to the successful assembly of the DNA-[2]-catenanes (v, ~23%). When taking all the assembly steps into account, the overall yields of strategies A and B are 4.8% and 23%, respectively. Strategy B achieves a higher yield than strategy A also due to the associated fewer assembly and purification steps (see Figure 3a). The enlarged TEM images of the interlocked half-ring complex and the DNA-[2]-catenane in Figure 3c clearly shows the presence of the interconnecting AuNP between the two origami components in the individual structure.

### 2.3. Unlocking the DNA-[2]-catenanes

To achieve the unlocked state, in which the two origami rings within a DNA-[2]-catenane can move freely relative to one another, the interconnecting AuNP is released from the structure by toehold-mediated strand displacement reactions as shown in **Figure 4a**.

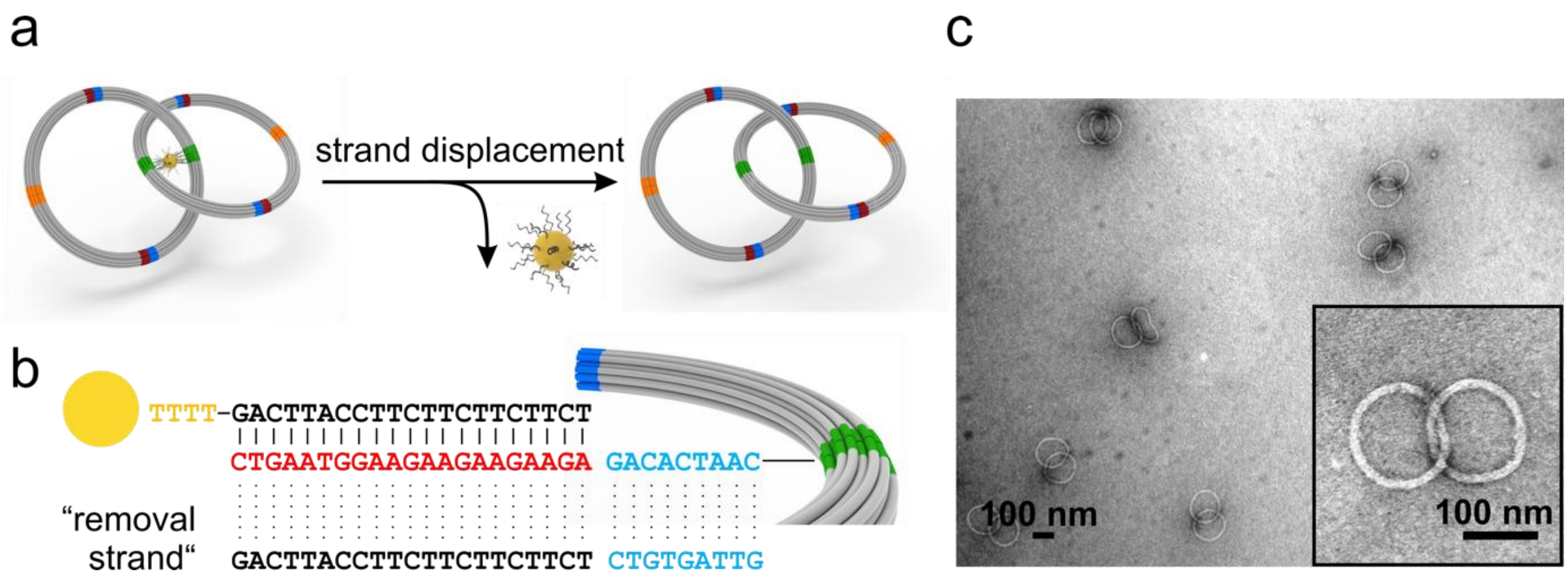

**Figure 4 | Unlocking the DNA-[2]-catenanes.** (**a**) Removal of the interconnecting AuNP by toehold-mediated strand displacement reactions. (**b**) Schematic of the AuNP-binding capture strand and the removal strand. (**c**) TEM image of the unlocked DNA-[2]-catenanes. Inset: enlarged view of a DNA-[2]-catenane after removal of the interconnecting AuNP.

Each capture strand extended from the origami contains two functional segments (see Figure 4b). One (red) is a 21-base segment for binding the AuNP and the other (blue) is a 9-base toehold segment. The removal strand completely hybridizes with the capture strand, releasing the interconnecting AuNP from the DNA-[2]-catenane. Figure 4c presents the TEM image of the DNA-[2]-catenanes after the dissociation of the AuNPs. The enlarged TEM image reveals the absence of the AuNP in between the two origami rings. When taking all the assembly steps of strategy B into account, an overall yield of 20% of the DNA-[2]-catenanes after the removal of the AuNPs has been achieved.

### 2.4. Assembly of the DNA-[3]-catenanes and DNA-[4]-catenanes

Taking a further step, we demonstrate the assembly of the DNA-[3]-catenanes and DNA-[4]-catenanes (**Figure 5**).

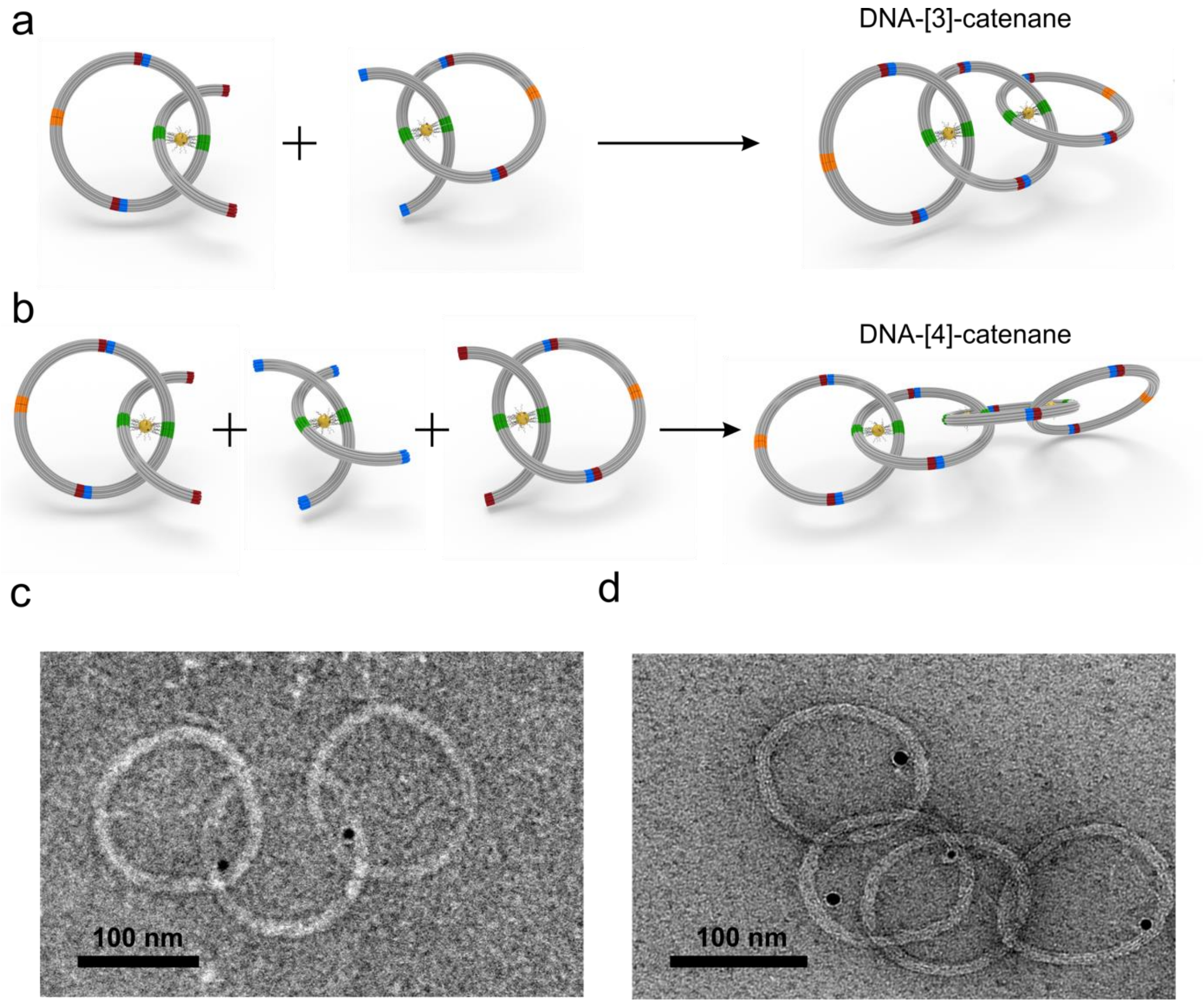

**Figure 5 | Assembly of the DNA-[3]-catenanes and DNA-[4]-catenanes.** (**a**) Schematic of the DNA-[3]-catenane formation. (**b**) Schematic of the DNA-[4]-catenane formation. (**c**) TEM image of a DNA-[3]-catenane with interconnecting AuNPs. (**d**) TEM image of a DNA-[4]-catenane with interconnecting AuNPs. Dislocation of the AuNPs is likely due to the distortion of the large three-dimensional structures, as a result of the dispersion and drying processes on the TEM grids.

To assemble the DNA-[3]-catenane, two different 1.5-ring complexes with junctions IIa and IIb at the two ends of the threading half-rings, respectively, are linked together as illustrated in Figure 5a. The DNA-[4]-catenane is assembled by linking two identical 1.5-ring complexes via an interlocked half-ring (see Figure 5b). The threading half-ring in each 1.5-ring complex contains junctions IIb at the two ends and the interlocked half-ring possesses the corresponding junctions IIa. Figures 5c and 5d present the TEM images of a DNA-[3]-catenane and a DNA-[4]-catenane with interconnecting AuNPs, respectively. Dislocation of the AuNPs is likely due to the distortion of the large three-dimensional structures, as a result of the dispersion and drying processes on the TEM grids. When taking all the assembly steps into account, the overall yields of the DNA-[3]-catenanes and DNA-[4]-catenanes are 5.3% and 1.7%, respectively.

## 2.5. Conclusion

We have demonstrated a AuNP-templated approach to hierarchically assemble DNA origami catenanes. Interlocked structures including DNA-[2]-catenanes, DNA-[3]-catenanes, and DNA-[4]-catenanes have been created by selectively connecting multiple origami subunits using AuNPs as mediators. Through toehold-mediated strand displacement reactions, the DNA origami catenanes can be unlocked after removal of the interconnecting AuNPs. Additional AuNPs could be assembled on the origami rings so that the removal of the interconnecting AuNPs could be optically monitored in real time as a result of the plasmonic coupling changes. Also, through smart designs controlled ring rotation within catenanes could be envisioned for the creation of DNA origami-based nanomachines and motors. This work will enrich the toolbox of DNA-based functional nanodevices and stride a step further towards advanced DNA architectures with programmable and well-controlled topologies.

**Supporting Information**

Supporting Information is available from the Wiley Online Library or from the author.

**Acknowledgements**

This project was supported by the European Research Council (ERC Dynamic Nano) grant. We thank Maximilian J. Urban for helpful discussions and Marion Kelsch for assistance with TEM. TEM images were collected at the Stuttgart Center for Electron Microscopy.

Received: ((will be filled in by the editorial staff))
Revised: ((will be filled in by the editorial staff))
Published online: ((will be filled in by the editorial staff))

**Table of contents**

**Andreas Peil, Pengfei Zhan, and Na Liu***

**DNA origami catenanes templated by gold nanoparticles**

Mechanically interlocked molecules have marked a breakthrough in the field of topological chemistry and boosted the fast development of molecular machinery. In this work, we demonstrate the hierarchically assembly of DNA origami catenanes templated by gold nanoparticles.

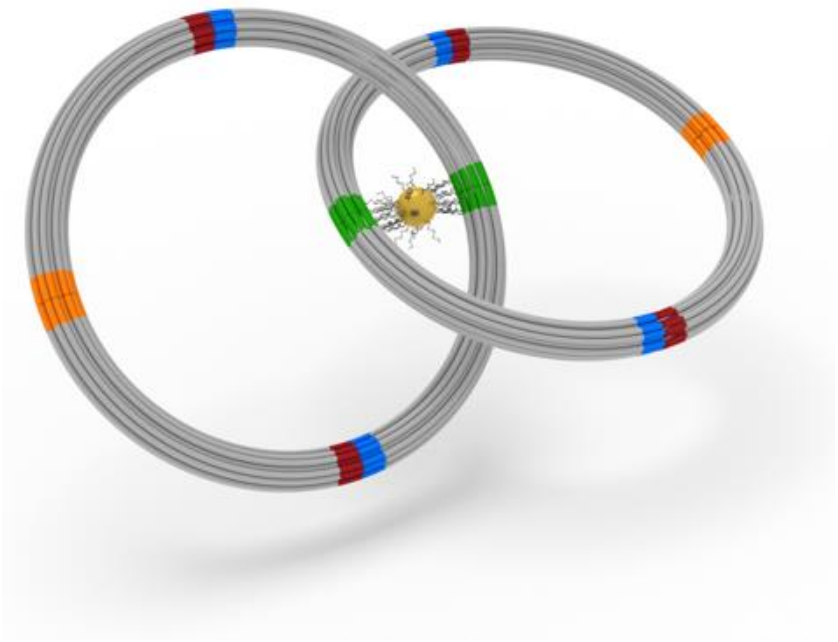

# Supporting Information

## DNA origami catenanes templated by gold nanoparticles

*Andreas Peil, Pengfei Zhan, and Na Liu**

**Content:**

## I. Materials and Methods

**DNA origami preparation.** DNA staple strands were purchased from Sigma-Aldrich (www.sigmaaldrich.com). The single stranded scaffold (p7560) was purchased from Eurofins Genomics (www.eurofins.de). The staple strands were split into 7 categories. They are the core mixture, core mixture for AuNP binding, connector strands I, connector strands IIa, connector strands IIb, connector strands III, and AuNP capture strands.

To assemble the origami half-rings, the staple strands were mixed with the p7560 scaffolds in a 1 × TE-$Mg^{2+}$ buffer (40 mM Tris, 2 mM EDTA, 20 mM $MgCl_2$, pH 8). The molar stoichiometric ratio between the core mixture staples and the scaffolds was 10:1, the ratio between the connector strand staples and the scaffolds was 20:1 and the ratio between the AuNP capture strands and the scaffolds was 40:1. The final concentration of the scaffolds was adjusted to 20 nM. The assembly mixture was then annealed in a thermal cycler at 65 °C for 10 min and cooled with a linear ramp from 60 °C to 40 °C in 21 hours. The temperature was set to 25 °C in a final step to store the assembled origami half-rings.

To assemble the origami half-rings incapable of AuNP binding, the core mixture, connector strands III, and connector strands IIa or IIb, respectively, were mixed. To assemble the origami half-rings capable of AuNP binding, the core mixture for AuNP-binding, connector strands I, AuNP capture strands, and connector strands IIa or IIb, respectively, were combined.

The DNA origami rings were assembled from purified half-rings. The half-rings modified with connector strands IIa and IIb, respectively, were mixed and assembled in a thermal cycler with a linear ramp from 40 °C to 25 °C (10 min/°C) for 10 cycles.

**Surfaces modification of AuNPs with BSPP.** AuNPs (15 nm) were purchased from Sigma-Aldrich (www.sigmaaldrich.com). BSPP (3.75 mg) was added to the colloidal gold solution (5 mL, OD~1) and the mixture was shaken over night at room temperature. Sodium chloride (solid) was slowly added to the solution until the solution color changed from deep burgundy to light purple. The resulting mixture was centrifuged at 3000 ×g for 30 min. The supernatant was discharged and the AuNPs were re-suspended in a 250 µL BSPP solution (2.5 mM). Upon mixing with 250 µL methanol, the mixture was again centrifuged at 3000 ×g for 30 min. The supernatant was removed and the AuNPs were re-suspended in a 250 µL BSPP solution (2.5 mM). The concentration of the AuNPs was estimated according to the optical absorption at 520 nm.

**Preparation of the AuNP-DNA conjugates.** The AuNP-DNA conjugation was done according to Kuzyk *et al*.[59] with minor modifications. The disulfide bond in the thiol-modified oligonucleotides was reduced using tris(2-carboxythyl)phosphine (TCEP, 100 mM, 1 h) in water. Reduced thiol-modified oligonucleotides and BSPP-coated AuNPs were then incubated at a molar ratio of 2000:1 in a 0.5 × TBE buffer (45 mM Tris, 45 mM Boric acid, 1 mM EDTA) for 1 h at room temperature. The NaCl concentration was slowly increased to 500 mM in the subsequent 5 h in order to increase the density of the thiolated DNA on AuNPs. The AuNP-DNA conjugates were then washed 5 times at 7000 ×g for 2 min using a 0.5 × TBE buffer in 100 kDa (molecular weight cut-off, MWCO) centrifuge filters to remove the extra free oligonucleotides. The concentration of the AuNP-DNA conjugates was estimated according to the optical absorption at 520 nm.

**Assembly of the AuNPs on the DNA origami.** The purified AuNPs were added to the DNA origami structures capable of AuNP binding in an excess of 5 AuNPs per binding site on the DNA origami structure. The mixture was annealed in a thermal cycler with a linear temperature ramp from 35 °C to 33 °C (2h/°C) followed by a second temperature ramp from 32 °C to 25 °C (4h/°C).

**Preparation of the interlocked structures.** To assemble the interlocked half-rings, the half-rings capable of AuNP binding and purified AuNPs were mixed in a molar ratio of 1 to 0.5. The mixture was then annealed in a thermal cycler with a linear temperature ramp from 35 °C to 33 °C (2h/°C) followed by a second temperature ramp from 32 °C to 25 °C (4h/°C). To assemble the interlocked 1.5-ring structure, purified AuNP-modified rings and half-rings were mixed in a molecular ratio of 1:3 and subsequently annealed using the aforementioned annealing program.

**AuNPs removal by the strand displacement reactions.** The removal strands (GTTAGTGTCTCTTCTTCTTCTTCCATTCAG) were added to the purified DNA origami catenanes in an excess of 10 removal strands per capture strand. After incubation for 24 h at room temperature, the AuNP clustering strand (CTGAATGGAAGAAGAAGAAGATTTAGA AGAAGAAGAAGGTAAGTC) was added in a molar ratio of 5:1 (clustering strand to removal strand). One clustering strand binds two free AuNP capture strands and thus cluster the AuNPs so that they can be removed by centrifugation. After incubation for 24 h at room temperature and centrifugation at 4000 ×g for 2 min, the free DNA origami catenanes were present in the supernatant.

**Agarose gel electrophoresis.** Annealed DNA origami structures and AuNP decorated origami structures were subjected to agarose gel electrophoresis for purification. Samples were run in a 0.7 % agarose gel in a 0.5 × TBE-$Mg^{2+}$ buffer (45 mM Tris, 45 mM Boric acid, 1 mM EDTA, 10 mM $Mg^{2+}$) at 80 V for 3 h within an ice-water bath. Sybr gold was used to stain the DNA origami structures. After segregating the gel bands, the DNA origami structures were extracted by squeezing and isolated using Quantum Prep™ Squeeze 'N Freeze filter units (Bio-Rad Laboratories, Inc., Hercules, USA) at 500 ×g for 1 min. The concentration of the purified DNA origami structures was estimated according to the optical absorption at 260 nm.

**Transmission electron microscopy.** DNA origami structures were imaged using a Philips CM 200 TEM operating at 200 kV. For imaging, negatively stained samples were prepared on freshly glow-discharged carbon/formvar FCF 400-CU TEM grids (Electron Microscopy Sciences, Hatfiled, USA). The sample solution was adsorbed on the grid and subsequently stained with a 2 % aqueous uranyl formate solution containing 17.5 mM sodium hydroxide. Samples were dried overnight. Uranyl formate for negative staining was purchased from Polysciences, Inc. Data processing was performed using the Fiji for ImageJ software.[60]

## II. Origami design

**Monomer design.** Monomer design of the 24-helix DNA bundle can be found in Figures S1 and S2. Constant bending of the DNA bundles was achieved by the uniformly deletion and addition of the base pairs along the helices. The DNA bundle was split into three parts, outer part, central part and inner part. Helices of the innermost part (helices 19 and 20, 1$^{st}$ layer) have 28 deletions, helices of the 2$^{nd}$ (helices 17, 18, 21, 22) and 3$^{rd}$ layer (helices 2, 3, 16, 23) have 21 and 15 deletions, respectively. Helices of the outer part have inversely 15, 21 and 28 additional base pairs, respectively.

The monomer was further split into two regions, a constant central region and two variable junction regions at the ends. The central region contains the constant part of the monomer. The base pair deletions and additions enabling the curvature of the monomer are located here. The junction regions are able to bind different staples. The region at the head of the monomer is able to bind connector strands I, the AuNP capture strands and connector strands III. The region at the tail of the monomer is able to bind connector strands IIa and IIb, respectively. The flexible junction regions allow the precise control of the connectivity of the monomers.

**Half-rings and rings.** Half-ring and ring structures were designed to form from monomers upon hybridization via short overhanging connector DNA strands. These connector strands contain two parts. One part anchors the connector strand onto the DNA origami structure and the second part forms the sticky end overhang which mediates the binding. Connector strands I and III contain 10 and 12 sticky end overhangs (3 bases), respectively. In total, 20 and 24 strands, respectively, participate in the head-to-head connection. Connector strands IIa contain 11 sticky ends (9 bases) but do not have the complementary free binding segments, whereas connector strands IIb have no sticky end overhangs but comprise the complementary free binding sequences. Thus, connector strands IIa can only mediate binding to connector strands IIb. Overall, 11 strands participate in the tail-to-tail connection in total.

## III. Supplemental Figures

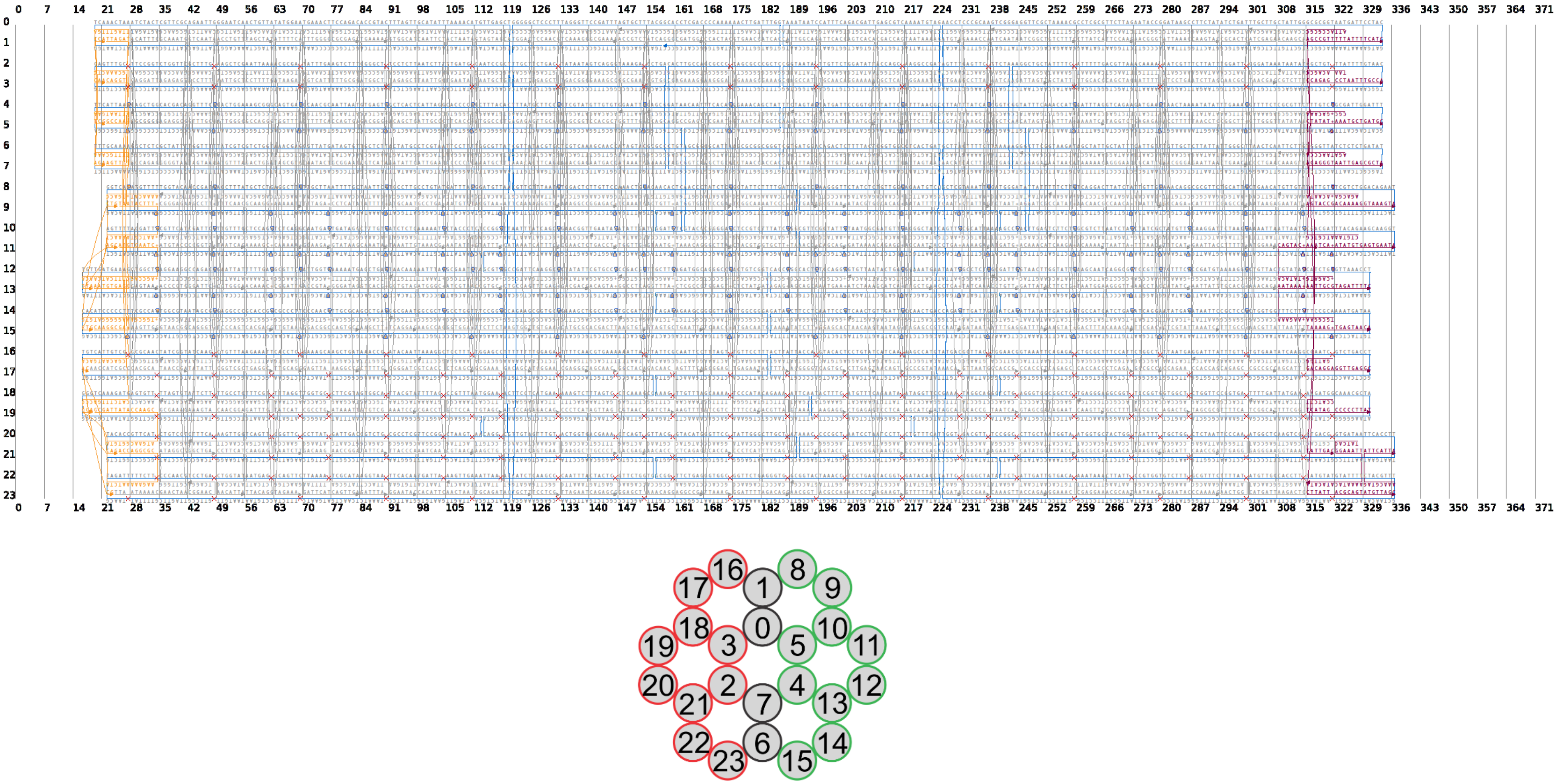

**Figure S1 | Cadnano[61] design and DNA bundle layout of the monomer modified with connectors IIb and III.** The DNA bundle layout shows the helices with base pair deletions (red) and base pair additions (green). Position and number of the base pair deletions (cross) and additions (loop) can be found in the cadnano design. The connector strands III (orange) and IIb (red) are located at the head and tail end of the monomer, respectively. Core staples are shown in grey.

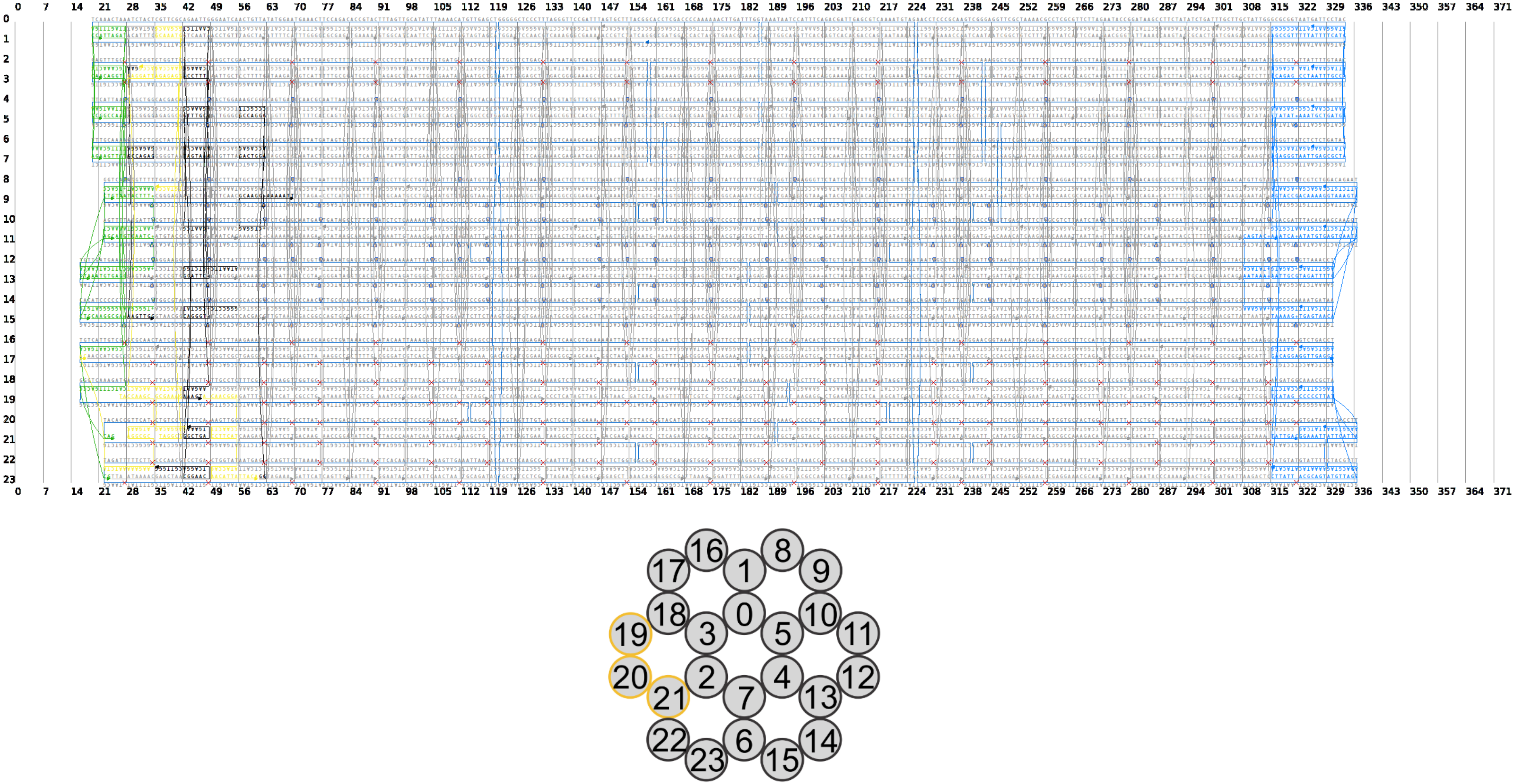

**Figure S2 | Cadnano design and DNA bundle layout of the monomer and connectors IIa, I and the AuNP capture strands.** The DNA bundle layout shows the helices with the AuNP capture strands (yellow). Positions of the AuNP capture strands can be found in the cadnano design (yellow). To design the AuNP capture strands, the core section of the monomer had to be slightly modified in the AuNP binding region. Required modifications of the core staples are shown in black. The connector strands I (green) and IIa (blue) are located at the head and tail end of the monomer, respectively. Core staples are shown in grey.

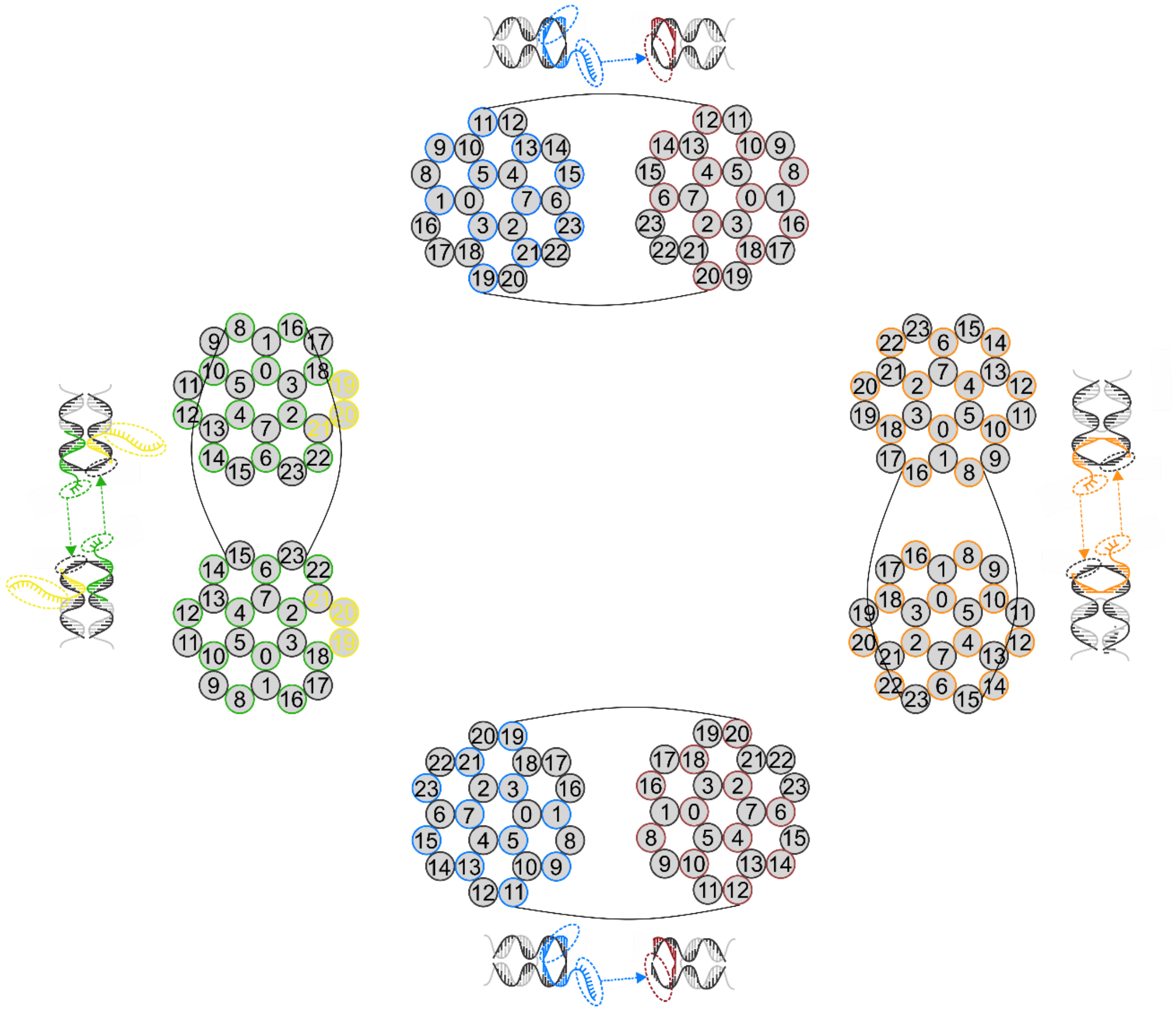

**Figure S3 | Schematic of the connector design.** The schematic illustration of the full ring structure which is reduced to the ends of the 24-helix bundle is shown. Two exemplary connector strands for the connection of two monomers are highlighted (black lines). The detailed connector design is shown next to the appropriate DNA bundle. Staples are shown in grey and connector strands are in color. Connector strands I are shown in green. Helices 19, 20 and 21 (yellow) harbor the AuNP capture strands. Connector strands IIa are shown in blue, IIb in red and III in orange. In total, 22, 24 and 11 strands of connectors I, III and II participate in the connection, respectively. Rotation of one origami by 180 degrees allows for the connection between two origami structures.

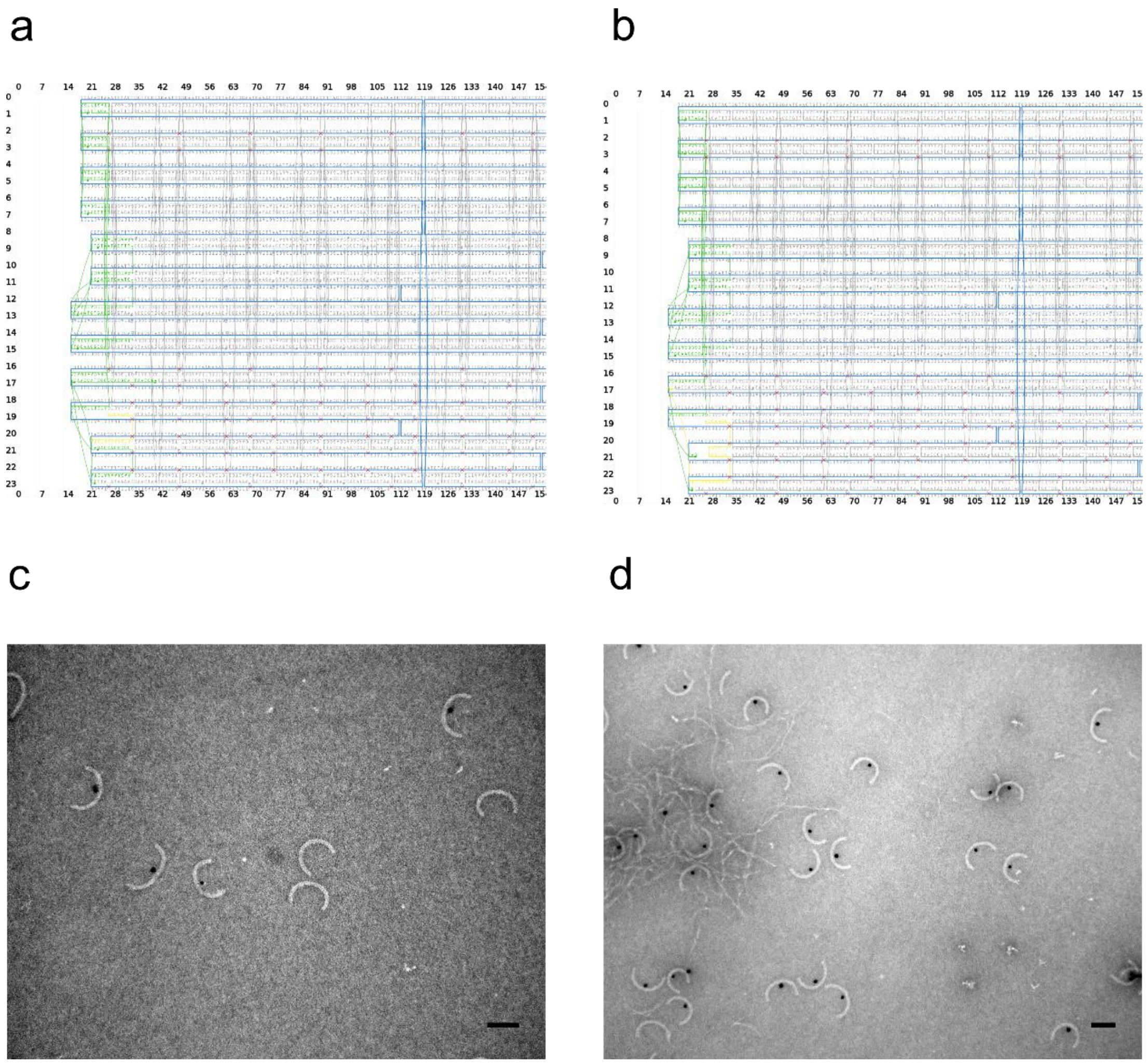

**Figure S4 | Influence of the lengths of the anchoring segment of the AuNP capture strand. (a)** Cadnano design of the AuNP capture strands with short anchoring segments (yellow). **(b)** Cadnano design of the AuNP capture strands with extended anchoring segments (yellow). Connector strands I are shown in green, staple strands in grey. **(c)** TEM image of the AuNP-modified half-rings containing the AuNP capture strands with short anchoring segments (scale bar: 100 nm). **(d)** TEM image of the AuNP-modified half-rings containing the AuNP capture strands with extended anchoring segments (scale bar: 100 nm). The TEM images show that the yield of the AuNP-modified origami half-rings is higher when using the extended anchoring segments. Almost all half-rings are modified with AuNPs. When using shorter anchoring segments, half-rings are only partly modified with AuNPs. The short anchoring segment may lead to dissociation of the AuNP capture strands due to lower binding forces resulting in half-rings without the AuNP capture strands and therefore unable to bind the AuNPs. Yields are shown in Table S1.

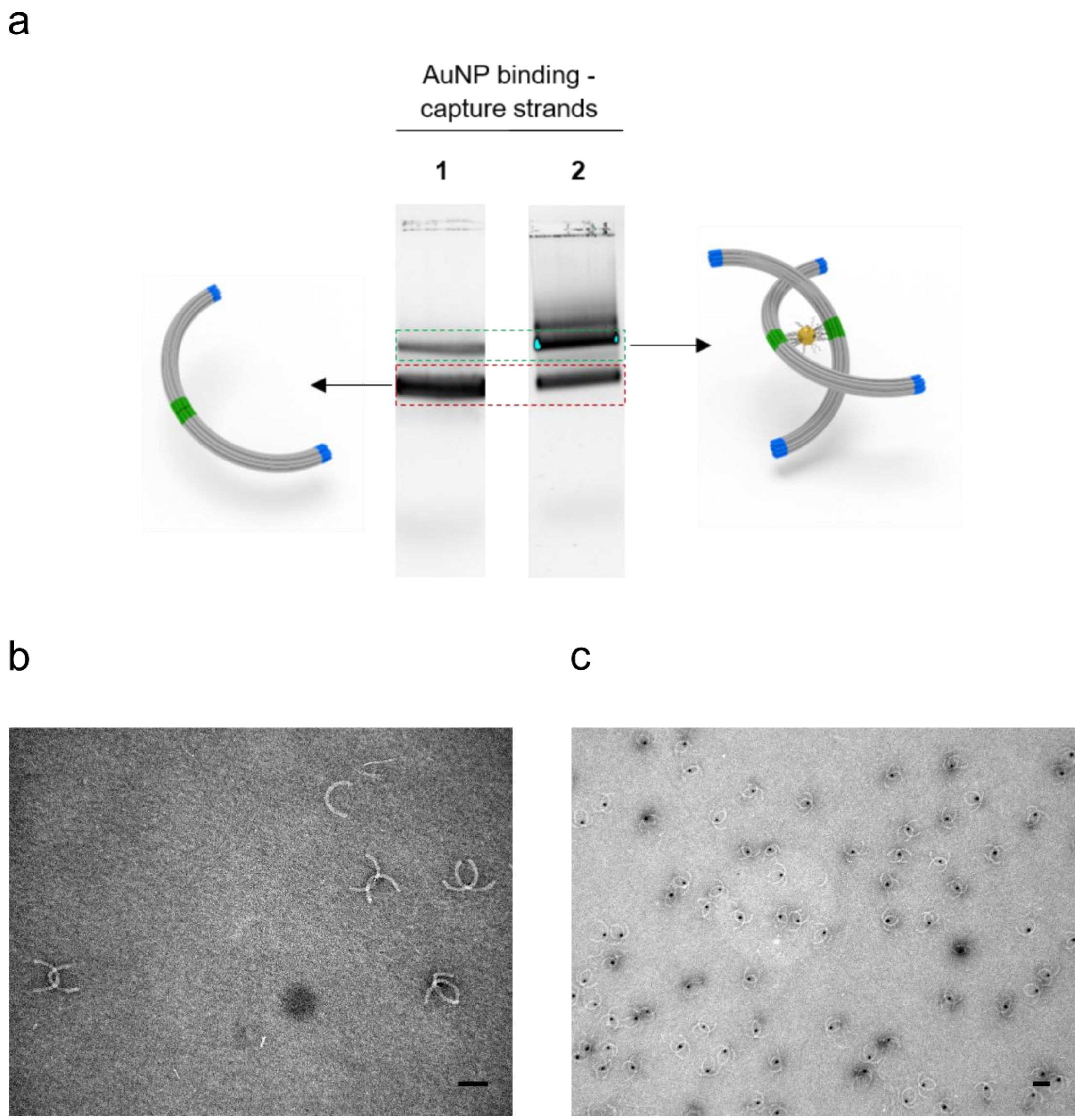

**Figure S5 | Influence of the AuNP capture strand number for the AuNP binding.** (**a**) Gel electrophoresis images of the interlocked half-ring assembly mixture. The ratio between the DNA origami and AuNPs is 1 to 0.5. The samples in the different lanes are as follows: (**1**) Half-rings modified with 2 AuNP capture strands. (**2**) Half-rings modified with 8 AuNP capture strands. The interlocked half-rings bands (green-dashed box) and the unreacted free half-rings (red-dashed box) are highlighted. **(b)** TEM image of the interlocked origami half-rings modified with 2 AuNP capture strands (scale bar: 100 nm). **(c)** TEM image of the interlocked origami half-rings modified with 8 AuNP capture strands (scale bar: 100 nm). The gel and TEM images show that half-rings containing 8 AuNP capture strands lead to higher yields of interlocked half-rings, indicated by the intense interlocked half-ring band of **2**. The TEM grid of the 8 AuNP capture strand sample shows a dense distribution of correct assembled interlocked half-rings, whereas the 2 AuNP capture strand sample only reveals a significantly lower density of correct assembled structures. Using only two AuNP capture strands, incorrect assembly or dissociation can lead to half-rings unable to bind AuNPs. The extended number of AuNP capture strands ensures that every half-ring contains AuNP capture strands. Moreover, the higher binding forces may guarantee more stable interlocked half-ring structures. Yields are shown in Table S1.

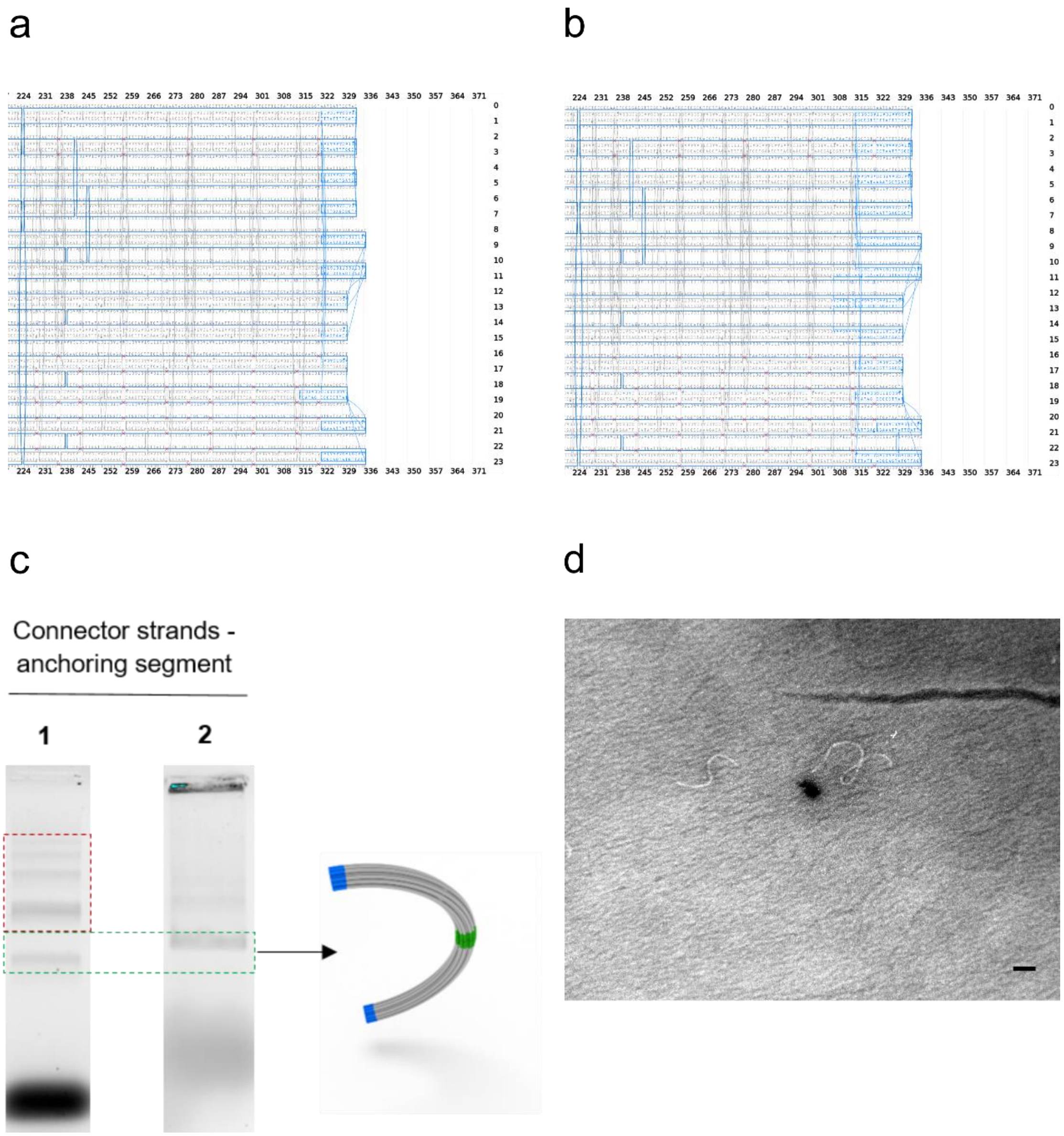

**Figure S6 | Influence of the lengths of the anchoring segment of connector strands II. (a)** Cadnano design of connector strands IIa with short anchoring segments (blue). **(b)** Cadnano design of connector strands IIa with extended anchoring segments (blue). Staple strands are shown in grey. **(c)** Gel electrophoresis images of the half-rings after assembly. The samples in the different lanes are as follows: (**1**) Half-rings modified with connector strands IIa containing the short anchoring segments. (**2**) Half-rings modified with connector strands IIa containing the extended anchoring segments. **(d)** TEM image of the side products of **1** (scale bar: 100 nm). The gel image reveals that many side products occur (red-dashed box) when using the short anchoring segments. Short anchoring segments accompanied by lower binding forces may lead to dissociation of connector strands what enables unspecific binding and therefore the formation of side products. The characteristic structures can be observed in the TEM image. Using extended anchoring segments prevents the connector strands from dissociation due to higher binding strength. This thus inhibits the formation of the side products and only give rise to the correctly assembled origami structures. Yields are shown in Table S1.

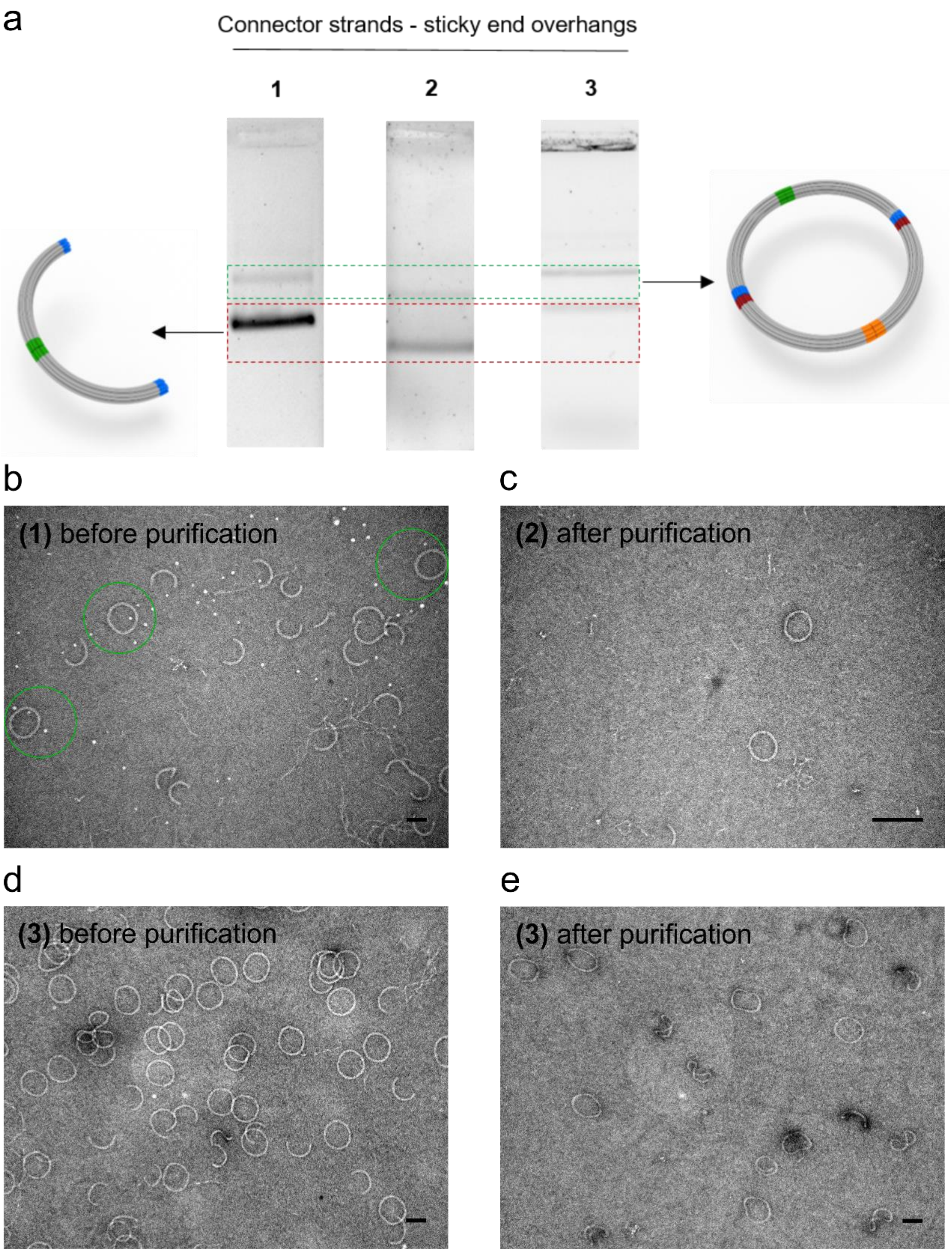

**Figure S7 | Influence of the length of the sticky end overhangs of connector strands IIa for the ring formation.** **(a)** Gel electrophoresis images of the ring formation. Half-rings are modified with connector strands IIa and IIb, respectively. The sticky end length of connector strands IIa varies. The samples in the different lanes are as follows: (**1**) 2-base long sticky ends. (**2**) 4-base long sticky ends. (**3**) 9-base long sticky ends. The origami ring bands (green-dashed box) and unreacted free half-rings (red-dashed box) are highlighted. **(b)** TEM image of the ring formation with 2-base long sticky ends before purification. **(c)** TEM image of the ring formation with 4-base long sticky ends after purification **(d)** TEM image of the ring formation with 9-base long sticky ends before purification and **(e)** after purification (scale bar: 100 nm). The gel and TEM images show that the 9-base long sticky ends lead to the highest yield of the origami rings. Almost all half-rings were assembled to rings, whereas many half-rings are not able to form rings when using 2- or 4- base long overhangs. When using short overhangs, the binding forces may be too low to generate stable origami rings. Yields are shown in Table S1.

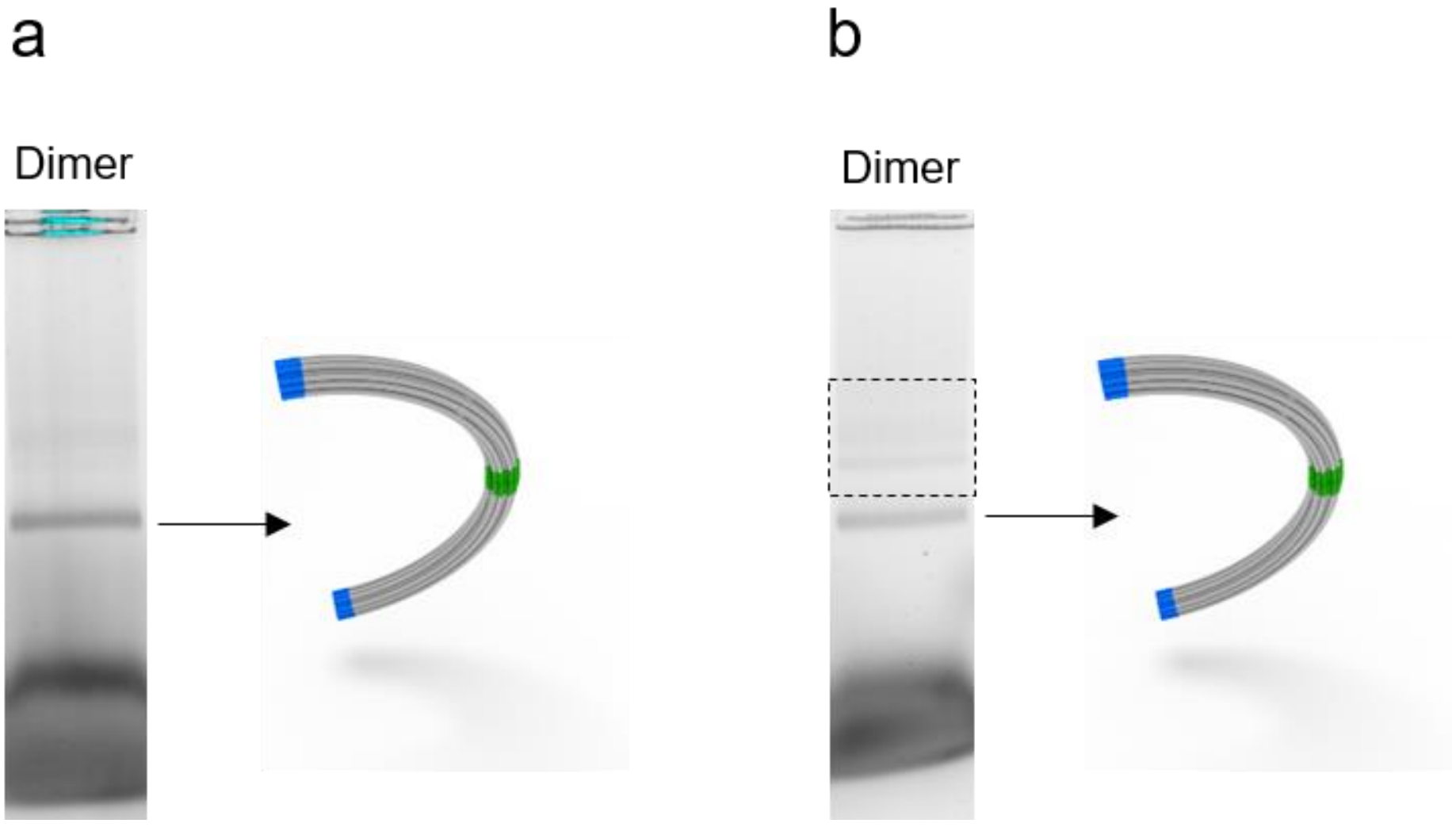

**Figure S8 | Influence of the assembly time and temperature on the half-rings formation.** Gel electrophoresis images of the origami half-ring purification after folding. **(a)** The folding program starts with an initial heating step at 65 °C followed by a temperature interval from 60 °C to 40 °C with a decrease rate of 1h/°C and one final step to 25°C. **(b)** The folding program starts with initial heating at 85°C for 5 min followed by a temperature interval from 70 °C to 26 °C with a decrease rate of 1h/°C and one final step to reach the storage temperature at 15 °C. The half-ring bands are labeled. The gel images show the formation of different side products (black-dashed box) when using the long annealing program. The short annealing program leads to higher yields and purer origami half-rings. The smaller temperature range and the fast temperature decrease may minimize unspecific binding events. Yields are shown in Table S1.

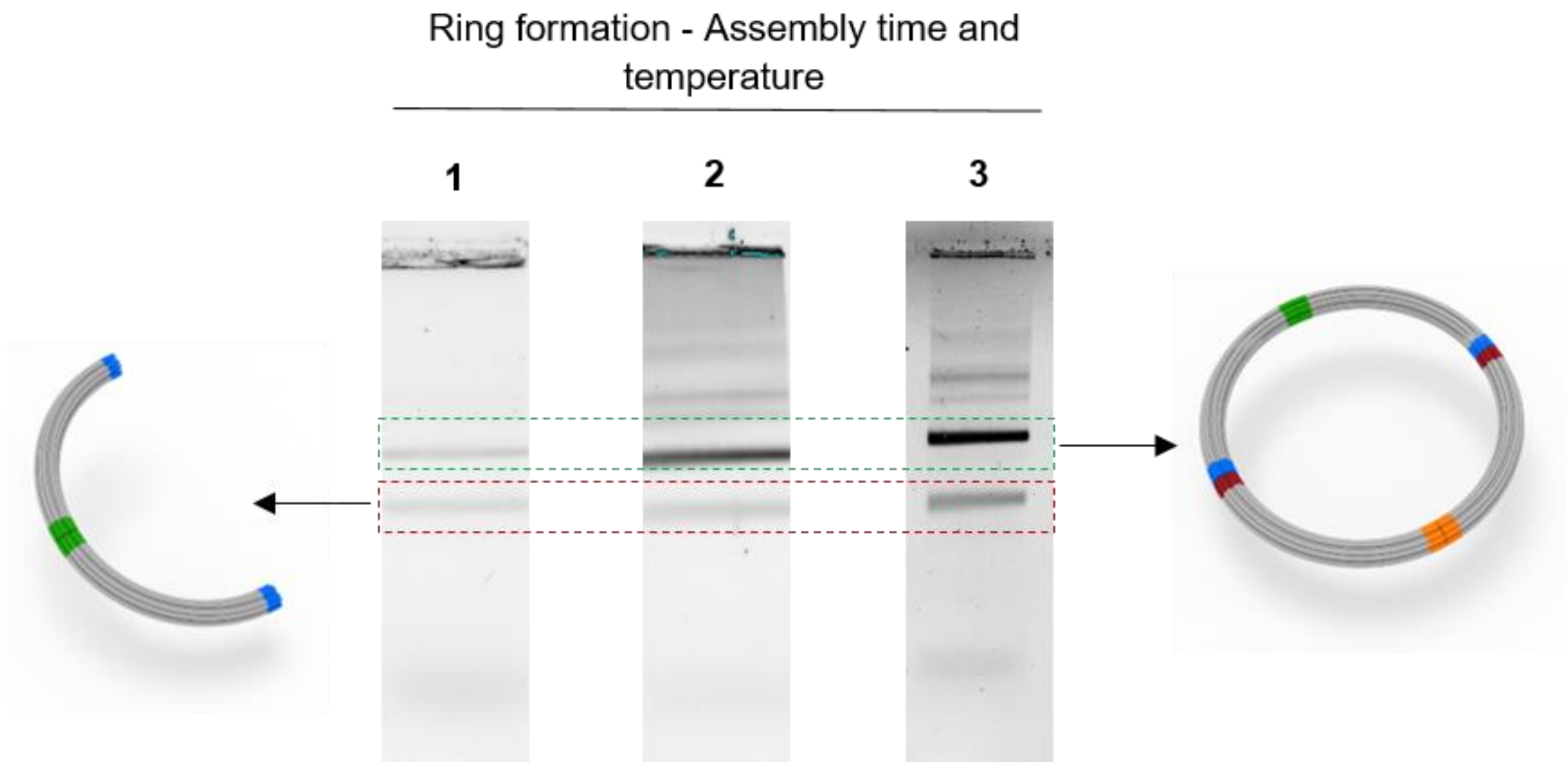

**Figure S9 | Influence of the assembly time and temperature on the ring formation.** Samples: 1:1-mixture of half-rings modified with connector strands IIa and IIb, respectively. The assembly times and temperatures in the different lanes are as follows: (**1**) Assembly at room temperature for 24 hours. (**2**) Annealing from 40 °C to 25 °C with a temperature decrease of 90 min/°C. (**3**) Annealing from 40 °C to 25 °C with a temperature decrease of 10 min/°C (10 cycles). The gel electrophoresis images show that **3** leads to the highest yield of origami rings, indicated by the intense origami ring band (green-dotted box). Annealing program **2** leads to a similar result. Most of the dimers are assembled to rings compared to **1**. Unreacted free half-ring bands are highlighted by the red-dashed box. The initial high annealing temperature results in higher meeting probabilities of half-rings. Lowering the temperature leads to DNA hybridization of the connector strands to the appropriate anchoring segments. The relatively fast temperature decrease may inhibit unspecific binding events. To further increase the yield, overall 10 cycles were applied. Yields are shown in Table S1.

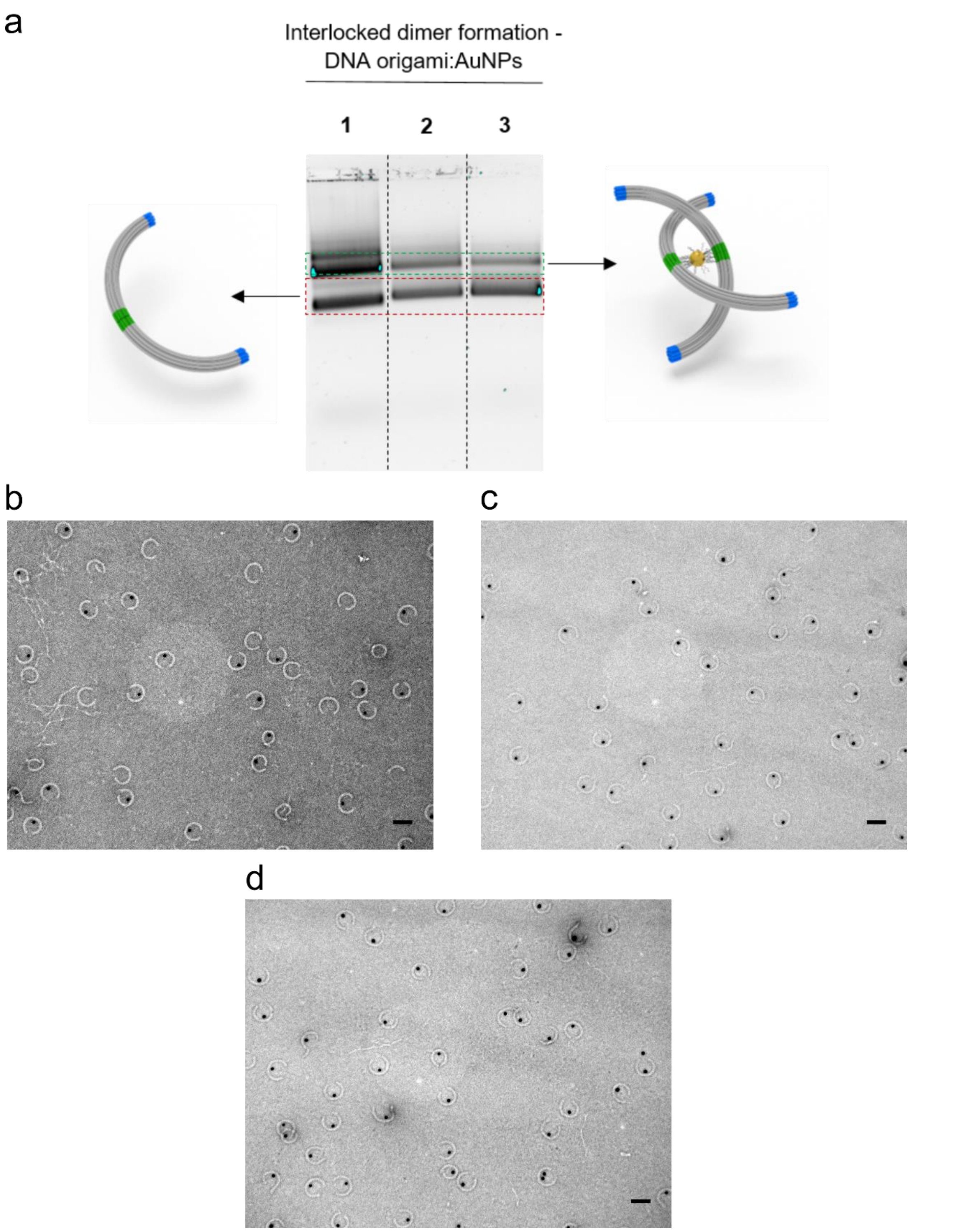

**Figure S10 | Influence of the DNA origami to AuNPs ratio for the interlocked half-ring formation. (a)** Gel electrophoresis images of the interlocked dimer assembly mixture. Each dimer contains 8 AuNP capture strands. The samples in the different lanes are as follows: (**1**) 1:0.5-mixture, (**2**) 1:1-mixture and (**3**) 1:2-mixture of half-rings to AuNPs. Half-rings contain connector strands I and AuNP capture strands. The gel images show that **1** leads to the highest yields of the interlocked origami half-rings, indicated by the most intense interlocked dimer band (green-dashed box). Unreacted free half-rings are highlighted by the red-dashed box. The TEM images show half-ring bands of **(b)** the 1:0.5-mixture, **(c)** the 1:1-mixture and **(d)** the 1:2-mixture (scale bar: 100 nm). The TEM image of **1** still reveals AuNP-modified and unmodified half-rings. At some point, the concentration of half-rings and AuNPs may be too low for the entwining event to occur. Furthermore, dissociation or blocking of the AuNP capture strands additionally may prevent the complete turnover. Higher AuNP to half-ring ratios (**2** and **3**) lead to AuNP-modification of almost all half-rings. Thus, entwining of two half-rings around one AuNP center is no longer possible. TEM image of the interlocked half-rings of **1** can be found in the main text (Figure 3c). Yields are shown in Table S1.

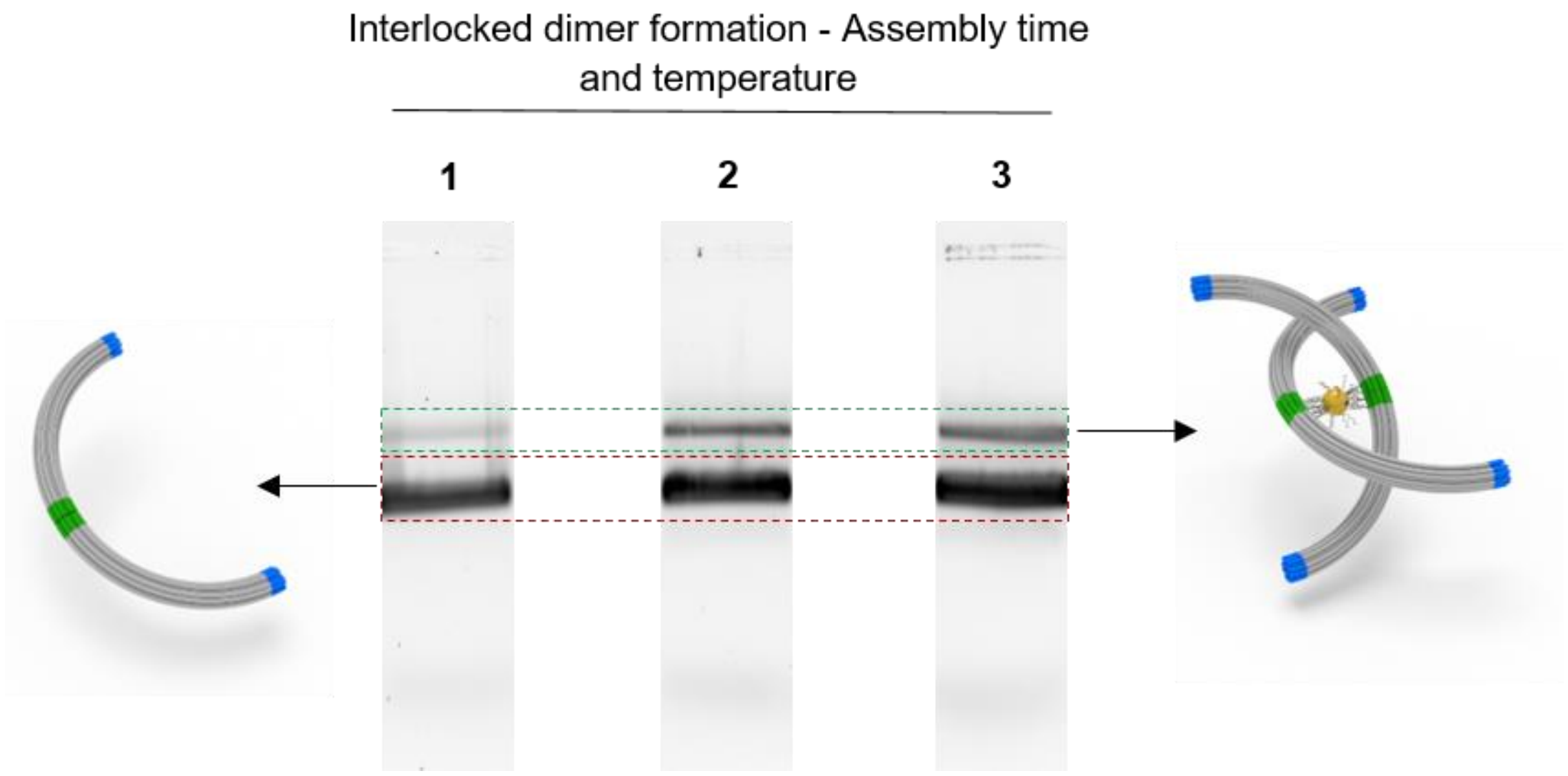

**Figure S11 | Influence of the assembly time and temperature for the interlocked half-rings formation.** Samples: Origami half-rings to AuNPs mixture in a ratio of 1 to 0.5. Half-rings contain connector strands I and AuNP capture strands. The assembly times and temperatures in the different lanes are described as follows: (**1**) Assembly at room temperature for 24 hours. (**2**) Four-step annealing program at 34 °C/3h, 30 °C/3h, 28 °C/3h and 26 °C/3h. (**3**) Annealing from 35 °C to 33 °C with a temperature decrease of 2h/°C followed by a temperature interval from 32 °C to 28 °C with a decrease of 4h/°C. The gel images show that **3** leads to the highest yield of the interlocked origami half-rings, indicated by the most intense interlocked half-ring band (green-dashed box). **2** shows a similar result, but the yield of **1** is significantly lower. Unreacted free half-ring (red-dashed box) is highlighted accordingly. The initial high annealing temperature results in higher meeting probabilities of AuNPs and AuNP-binding sites. Lowering the temperature then leads to the binding process through DNA hybridization. Yields are shown in Table S1.

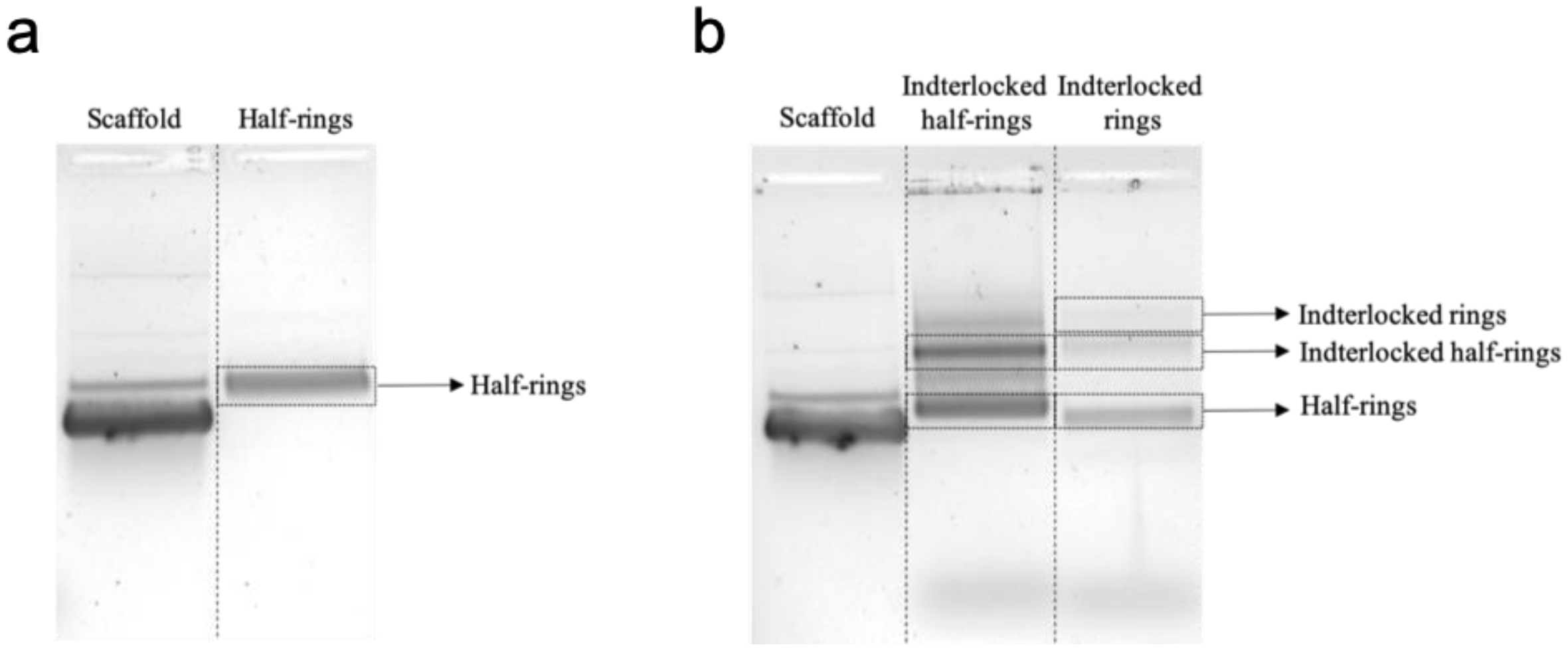

**Figure S12 | Agarose gel images for strategy B.** (a) Agarose gel image of the p7560 scaffold control and half-ring structures. (b) Agarose gel image of the p7560 scaffold control, interlocked half-rings, and interlocked rings.

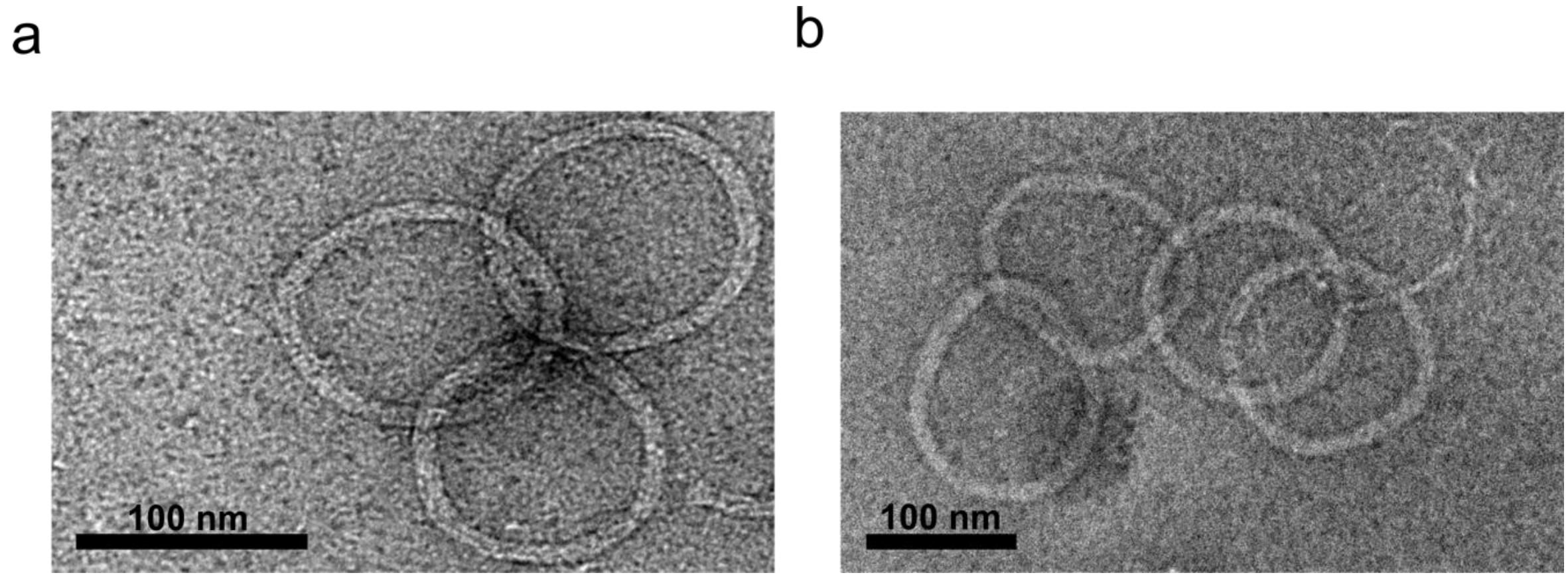

**Figure S13 | TEM images of DNA-[3]- and DNA-[4]-catenanes after the removal of the interconnecting AuNPs.** (**a**) TEM image of a DNA-[3]-catenane after removing the interconnecting AuNPs. (**b**) TEM image of a DNA-[4]-catenane after removing the interconnecting AuNPs.

**Table S1 | Yield chart of the assembly steps.** For Figure S4, a statistical analysis was done by counting the correct and false structures on the TEM grid. The yields of Figures S5 to S11 were determined by gel analysis using the ImageJ and Magicplot (©2016, Magicplot Systems, LLC) software. Analysis was done by plotting the relative density of the gel lanes. Each peak of the plot represents a DNA origami band in the gel. After peak fitting followed by calculation of the peak integrals, the product yield was determined.

| | **AuNP-capture strands** | **Yield [%] of AuNP-modified dimers** |
|---|---|---|
| Figure S4 | Short anchoring segment | 51 |
| | Long anchoring segment | 87 |

| | **Number of AuNP capture strands** | **Yield [%] of interlocked dimers** |
|---|---|---|
| Figure S5 | 2 | 27 |
| | 8 | 45 |

| | **Connector anchoring segment** | **Yield [%] of dimers** |
|---|---|---|
| Figure S6 | Short anchoring segment | 21 |
| | Long anchoring segment | 66 |

| | **Connector overhangs** | **Yield [%] of origami rings** |
|---|---|---|
| Figure S7 | 2-base long sticky ends | 12 |
| | 4-base long sticky ends | 16 |
| | 9-base long sticky ends | 46 |

| | **Annealing program** | **Yield [%] of dimers** |
|---|---|---|
| Figure S8 | (a) | 73 |
| | (b) | 67 |

| | **Annealing program** | **Yield [%] of origami rings** |
|---|---|---|
| Figure S9 | (1) | 46 |
| | (2) | 62 |
| | (3) | 62 |

| | **Ratio origami:AuNPs** | **Yield [%] of interlocked dimers** |
|---|---|---|
| Figure S10 | 1:0.5 | 50 |
| | 1:1 | 45 |
| | 1:2 | 25 |

| | **Annealing program** | **Yield [%] of interlocked dimers** |
|---|---|---|
| Figure S11 | (1) | 10 |
| | (2) | 23 |
| | (3) | 24 |

## IV. DNA Sequences

**Table S2 | DNA sequences of the core staples of the monomer.**

| | |
|---|---|
| Core-1 | AACACTCAGGATTACCGGAAGCCAGAGGAGAGGCAAGTTGG |
| Core-2 | CCAGAAAGCCACATCCTGAAACGATTGACGGGAAGCAAGACTCGTATTAAATCC |
| Core-3 | GCGAAATTCCATTAATGCTCGGATTGAAATGCTAACGCCATTCTTCTAAGTGGT |
| Core-4 | CGGGATCTACGTAATAGAGCTAAGAGGAAATATTCGCATAGTTCAGGAG |
| Core-5 | ATGTTTAGACGATATCCCAGTCACGACG |
| Core-6 | TGAAATACATAGTGGTCAAGTAGAATCGCCATATTTACAAGAAA |
| Core-7 | CGTCTTTCCAATAGACAAGCAAACAACA |
| Core-8 | CGCTCAACCGACCGCCTGTTTG |
| Core-9 | AATTTATCCGTTTTAGCTATTGAGCCGCAGAATCAAGTTT |
| Core-10 | GTTGGGTAGAAAACAAACAGA |
| Core-11 | ACACTATGCGGAATATTAATTCCAGTGAGCTATCA |
| Core-12 | TTGCACCCAAAAATTAAAAACAATAGCTGACTTTACAAACAA |
| Core-13 | GGCGAACTTTGAGCGTAACGATCTA |
| Core-14 | GGTTGCTAATGACCACACAACCCACGCTGATATTCA |
| Core-15 | GTACTATGGAGCTCCAGAACGAGTAGTGCCCT |
| Core-16 | TTATCCCCCCTGAATGAGTTAAACGTTATTAATTT |
| Core-17 | CTAAATTTAATGGAGATGGCAATTCATCAATATAATGATTATA |
| Core-18 | AGTAAAAGCAATACACACCGGATATGCG |
| Core-19 | TAAGCATATAACATATAAGGCCCAGTATCGAACTGATTGAATGGCTATTAG |
| Core-20 | CTTTTTAATTTGAGGATTTAG |
| Core-21 | AATCCAAATAAGAATCTTACATAGAAGCAAGTAC |
| Core-22 | GTGGCATAACTAAAACAGCTGAGTGAGCGGTGTAGATTCGCC |
| Core-23 | CCCTGACGAAAGCCAAATCGGGCGAAAA |
| Core-24 | GCCACCCTCAGAACTCAGAGTGGCAAGAGAAAGGGTCGAGGCGATGGC |
| Core-25 | CGACCCGACCACCAAAATAGCTGGTCAATCGATAAAAATCCCTTATTGAAA |
| Core-26 | GTTCAGAGTCAGAAGTGAAACAAACTACAACGATTCCACAGGAAGTGACAGCA |
| Core-27 | ATAAATCTCTTTAGACGAGA |
| Core-28 | AATTAACTATTACGGTGCTC |
| Core-29 | CTTATCACGCGAGGAAAATCA |
| Core-30 | TTACGAGATTGAATCCTAATGATTGCCCCTATTTTTGCCTGAGT |
| Core-31 | TAACGATTTAAACAATAAAGTTGAGAGAAAATTAAATTCAAAAGGGTGAG |
| Core-32 | CCGCTACGCGGTCAATTGTTAGCCGAGCGGTACGAATGGTGGTTCCGAAA |
| Core-33 | TAAAGCCTATTTTCAGAGGTGTTGCAGG |
| Core-34 | GCAAAATTAAGCAATTTAGAACGGTCATTGGGAAGATT |
| Core-35 | CAAGGCAATGCAATGAGAGATTAAATTGTTTGTTA |
| Core-36 | GAACGTTAACATCCAGGTAAAGTGCCGGAGGTATTTCA |
| Core-37 | CTATTAAATCCAACGTCTCCATCGGAACGAGGGT |
| Core-38 | GGAACAAGAGTCCAAAAGGCCGGAGACAG |
| Core-39 | TAATAAGGAATTAAATATATTTAACCTCATCCTTGCATTTTCGAGCCAGT |
| Core-40 | CCGAGATACCACTACACTAAAGCGGAGTGAGAAT |
| Core-41 | ATCAAAAGAATAGCTCGGCAAGACGGAGCTACGTG |
| Core-42 | CTGACCAATTGGCAGTGGTAAACGGGGTCAGTGC |
| Core-43 | TGCCGGAAACCAGGCGGTGGATGAAAGGAA |
| Core-44 | AGACTTTTGACGGGTATTATAAAACGAGTTGACGACATGCGCACGACTTAAGTGTCC |
| Core-45 | GCGAACCATCGGCTGTCTTTC |
| Core-46 | CTAATTTAGAAACCAGTAAGCGTTAATGCCACCG |
| Core-47 | ATCAACAAAACGGGTAGCCAGGCCGCCA |
| Core-48 | GAACGCGCCTGTTTGGCAGAGGAAAACATATTTCATTT |
| Core-49 | CTAATGCACGCACTCAACAAACAGAGCCGCCGCC |
| Core-50 | TGTTCAGAATAAGATAATTTTATGGAAA |
| Core-51 | AAATAGCAGGGAAGTTCTGACTAGGTCTCTTAGATTCCAACATGTAATTTA |
| Core-52 | GACGCTCCGCTCATATTCTTCCTCAAGCATCG |
| Core-53 | CGGGAGAAGCCTTTAAAATCGGTGTCAATACTTGATACCGATA |
| Core-54 | CCTCATATATTTTAAAAGAATTATGAAAAG |
| Core-55 | AATGTGTAATAAATCACTAATAAAAGGACTCAGCA |
| Core-56 | TCAAATCTCCAGTTTACCGTCTTAATTTGGCTACA |
| Core-57 | ATACGTGGCACAGACGGCCAACACGACCAGTTGATGAACAGT |
| Core-58 | CGTTGTAGAGTCTGGGAGCACTAACAACCAGTTGA |
| Core-59 | GAGAATCGGTTTGCGTCAATTCTACCTTTAAGAAT |
| Core-60 | CCTGAGATGGTTTTTTTCCATAATAAGAACCTAAATATCATCGCCTG |
| Core-61 | CTACAAAGGACGGGCAGTACGGTATGGCTTGCCACGTGTCGAAATCCG |
| Core-62 | TAGCTGATGTTGCAGCAAGGGAGAGAGCTTTCATGAGACAG |
| Core-63 | ACCGTTCCGAACTCTGACCTCC |
| Core-64 | GATCCCCGTCGAATTCTTTTGGGAAGGGATTTTGCGTCGTCTTTCCAG |
| Core-65 | GAACGCCAGCCTGTTGATTATTAGCCATATTGAATGTAAATAAGAGG |
| Core-66 | GCCATTAAAAATACCCAGCAGAA |
| Core-67 | TAATGCGAAAGCCAATTGCGGGGCCTTACTGCCTAGCACCGTAATCA |
| Core-68 | GCGATAGGAGAGACTCGGTATTAATTTTCACCGGAAGCGTCAGACTG |
| Core-69 | CCCTTAGACGGCTTAGATCAGATCAACGCTAACGAG |
| Core-70 | CAAAGTCGAACGCGTATATAACTATTAATGAATATAA |
| Core-71 | CAGAGCATAAAGCTTTTCAACCAAACAATAATCAG |
| Core-72 | AATATTTTAGGGTAGTTCACCGCTGTTTTATGAATATAAACGGGCTCCATGTAGC |
| Core-73 | TGGTTGGTTCAATATGGTTTGCCAAGCACT |
| Core-74 | GATAAAACAGAGGTGAAAACATCTTATACAA |
| Core-75 | GGGTTGAGTGTTGTACTGTTTGAAATCCCAGTAATGAGGTAAGCAACTCGTCGG |
| Core-76 | ATAATTCGCGTCTGGGGATTCTCGGGAAACCTAGTAAACAAAGCGGCTGA |
| Core-77 | TAACCAATAGGAACGCCGGATTGGCTCACTGTAGCGTCTCGCGTTAATCTT |
| Core-78 | AATCAGCTGTCACGTTTAACTCACCGTCATAAGCCCGTATTCA |
| Core-79 | TGACGCTAAAATTCGCATCGTAACTGGGGTGCCCCCTCCATCAACGTAACA |
| Core-80 | TGGGCACGAATATAGGGGCCTTGTTTGAGGGCCGGAAGC |
| Core-81 | CGACAGTGCGGCCCTTGGAGTGATGTGTGAACGCTGCGGGGCGCCCACCA |
| Core-82 | CGCCTGCAACAGTGCCAGCAGCAATTACTAG |
| Core-83 | CATTTGCGGTCAGTAGCATCACCAAGAATAATTCTTTGGCTGGTTAAGTG |
| Core-84 | GCAGAGGCTATCAAATGTGATAACACTTGCGAACTCTGATATA |
| Core-85 | CGAGCGGACATTCTAATATTTTAGCCCTAGCAATT |
| Core-86 | AATAATATCCCATCTCTAATTGAGGGCTTTATGATGAAAAAATCGC |
| Core-87 | GAGAAACAATAACGGAAACCTACATTTCATCCGCATTATTTTGTAAAGGGC |
| Core-88 | GTAACAGTACCTTTTTGCACGTATTTTTCAACTGAACA |
| Core-89 | AATGAAAAATCTAAATTAACAC |
| Core-90 | TAGATAAGTCCTGAAACAACGAAGACGCACAAAAT |
| Core-91 | AATCGTCAGTCACAGAGATAGAACCCTTGCGTAAGA |
| Core-92 | ACTGAGTTTCGTAGTTAGGACTAAAGCAAC |
| Core-93 | CCTGTTAAGCTGTTAATTTGCAGATACTGCTT |
| Core-94 | GGGCCTCTTCGCTATCAGGGTTTAAAACCACGGAACTCAGGA |
| Core-95 | ATCGGTGCACAAACGGCATCAAAAAAAAGCCCCAAAAACA |
| Core-96 | CTGTTGGGAAGGGCGTTGTAAAACTCGTTTTAGAAA |
| Core-97 | TGCGCAAGGGATAGCATTTTTGTATAAGCAAATATT |
| Core-98 | ATTCAGGCAAGCTTTCAAGAGCAGATTTA |
| Core-99 | TAATCAAATCAAAAAGATTTAATTGCAATATGCCAATTCTATACAGG |
| Core-100 | CTTCTGGCTGCCAGAATCGGCCATAAATCATTTCTC |
| Core-101 | AGCTTTCCGGCACCGTGTGAATTGCACGTATCCTCG |
| Core-102 | CTCCAGCCACAGTATCGGCCTCAACAATTCC |
| Core-103 | CGCTTCTCTCGCCCGCCATCTTAAACAGGGCTTAAG |
| Core-104 | CCGCCAAAATAACCCTCAACCTTTAATGCGTTAAAG |
| Core-105 | GTTATCTCATAGAGCCACGCTGACGTGCTTGTTAAACCAC |
| Core-106 | AAGGAATTGAGGAAGTATCTTTATCCATCAAACGGTGTACCGC |
| Core-107 | CAGTTGGCAAATCAATAATAGATGCCACCGAGAAGT |
| Core-108 | TATCTGGTACCTCAAAGAATTATTACCTGAGCAAAAGAAG |
| Core-109 | CCTGATCTCAATCAAGATAATACAGAAAAGGCCGAA |
| Core-110 | TATCAGATAGGGTTAGTTCGCCTGTAATTACATTTAACAA |
| Core-111 | CATCATATTCCTGATTTCGACAAAACAATGAATAAC |
| Core-112 | GAATTATAATTATTACATCGGGAATTACCTTTTTTA |
| Core-113 | CCACCAGAAGGAGCGTTTGCCCGAGCCCAAAACTGGATAAAAGAAACGCCAACCGA |
| Core-114 | ATAAGGGACAAGTTTTAAGATTCATCAGTTGA |
| Core-115 | GAACCGAACTGAGATTTGACGAAATGAGGC |
| Core-116 | GTAACGCTACGCCAACCCGTCCCTTCCTATGTACCCCGGTTGA |
| Core-117 | AAGCCAGAAAGCGCCATGGGCGCATTAAATTTAAACGTT |
| Core-118 | AAGTATTAATCTTACCGAAGCC |
| Core-119 | CAGACGGTCAATCATAAATTTACGAAGAAGGCCG |
| Core-120 | CGAGGCGTTACCCATTGTGAAGGAATAC |
| Core-121 | AATTTCTTAAACAGACCTGTTTGATTCCATTGGGCTCCAGTCCGTGGGA |
| Core-122 | ATCAGCTTGCTTTCTTTGGGGAGTTTCACTTTTCAGCGTTGCACCGTAAT |
| Core-123 | ATTGTATCGGTTTCTTTTG |
| Core-124 | AAAAAAAGGCTCCAGTAGTAGCTCAACACTGGCCCGTAAAGCCCGTGCAT |
| Core-125 | TTTCACGTTGAAAATCAAAGGAACCCTAAAGCGGTATACGAGGACGACG |
| Core-126 | GGAATTGCGAATAAATCAGGGTGCCGTACCAGCAGTCCGCTC |
| Core-127 | TAAGTAAAGGAACAGTGAACCTCAAGTTGTAATCATGTTTCCCTCTATG |
| Core-128 | ATACAGGAGTGTACATTCACCTGAAATGTAGTATCAATCATA |
| Core-129 | GTCATACATGGCTTTAATAAAACATTTTATTCTTAGTTAAATTTGCTGA |
| Core-130 | AATTTACCGTTCCAATCAATATCCCGAC |
| Core-131 | AATGGAAAGCGCAGTCTCTGCAGAGCCACCACCTCCCTCA |
| Core-132 | TCATTAAATTAAACGCTTATCACCTTTTTTAGTTACATATCAA |
| Core-133 | CCCAATACCCTCATCAGAACCGGATTTTAGACAGG |
| Core-134 | AACCATCGCCCACGCATAACCGA |
| Core-135 | TATATTCGGTCGCGAGGCAAAATTGCTCGAGCTT |
| Core-136 | GAGTTAGCACCAGGTCATTTCAAATACAATACTCATAACCCGACGGCCAGTGCC |
| Core-137 | GGCTTGAGATGGTCTCATTCAGCAAAGGTGATTTCCCCCAGCATGGAC |
| Core-138 | TAGCCCGGAATAGGAGGGTAAACTATACTTGAAAGGTTCTAGATGTA |
| Core-139 | GACAACCTTTGGCGTGGCGTGTCAGGAAAAATAAGGGCGCTTAGTGCTGAATTG |
| Core-140 | GCCCGTAAACCTATTGGAAATCGGCCTTATTAGTACAGTGAGTAGAGCCGTCAATA |
| Core-141 | AAGTTACACAAACATCAAGAAATGAGAAGA |
| Core-142 | AGCATTCCTTGATATTCACAATCGAGACAAGCAA |
| Core-143 | AACGTCACCAATATAGCATTTCGGTAAACA |
| Core-144 | GCCGGAAGAATTAGAATACAAAGTTACCAGAA |
| Core-145 | TATGGGAAGAAAGCGGCGCTACGTAACCCCGCGCTATGACAATGTAAAA |
| Core-146 | GAAAGTTGCAACAAGAACAA |
| Core-147 | CAGACCAATCAAGTTTAGAACTGAGTACAGAGAG |
| Core-148 | TCACCGTTTTCATCGGCAT |
| Core-149 | GCACCATAGCGCCATTTATTTCGAGGAAACGCAAT |
| Core-150 | AAATCACCAGTAGCCTTTACCGCCCTCAGA |
| Core-151 | AAAGTACAACGGACCAACTTTGAAAG |
| Core-152 | CCCTCATCACCAGTTAAGGCTTAAATTGTTAGAATCAGAGCG |
| Core-153 | AGGACAGGCGAAACACACTAA |
| Core-154 | CACATTCAACTAATCAACTT |
| Core-155 | GGAAACTGTCACAATCAATCATATGGAGCAGCCATTAGTT |
| Core-156 | CGTCACCGACTTTTTCGGTTCATAATCAAAA |
| Core-157 | GAGCCATGAGGGAAGCAACATCATGATTAAGACTCCAAGAAT |
| Core-158 | ACTTAGCCGGAACGACCTGTAAAAGTCACCGCCTTTA |
| Core-159 | GGTTTACGAATCTAAAGATCGCA |
| Core-160 | ATTGCTTTGAATACCCTTCTGAATAATGGTTAATAACAGAAAATTTTACC |
| Core-161 | CGTAACAACACCCCGCCAAAACAGGAGGCCGA |
| Core-162 | GGAACCCATGTACAAGTTTTTAAACAAAGTTTCA |
| Core-163 | AAGGTGGGTAAAATATTAT |
| Core-164 | GTTTATAGCAAGACGTTATTTCTGAGTTTA |
| Core-165 | AGCGGGAGTACCGGTTTAACGCCAGAATCCTG |
| Core-166 | AAGGATTAGGATTCTGAGACGAAACATCTTGAGT |
| Core-167 | GAAACAGCCGTCGAGTGTATCGTTTTTATAAGATA |
| Core-168 | CCTTCATATACCAGAACATTA |
| Core-169 | TTGAGGTTGGGAATTAGAGTAGCGCGGAACCAGCCACCACTAAATCC |
| Core-170 | TACCATTAGCAAGGTAGCGACCACCCTCCACCCT |
| Core-171 | CGTTGGGATAGGCTGAACCAGAGAGAGTGCGAACGGCAAATGTGTACCA |
| Core-172 | ACCGTACTCAGGAAGGCGGAAATATCCGGAAAAA |
| Core-173 | CCAGCAGACATTAAAGACGGAATACCCAAAAG |
| Core-174 | TTAATAAAACGAACTAAAAATAGCGGGGTAATGTCGTGAGAGGCGATGAACG |
| Core-175 | TTACAGGACCAGACGACTGGACCCGCTTGCCAGGGGTCTGGAGGCAAGGATAAAAATT |
| Core-176 | ACCACGGAATAAGAAGACATTAACGTCAGCTACCTAAGAATTCCAAG |
| Core-177 | GAACTGGCTCATTCAAGAGTTTAATTCCTTTTGTAACAGTTAGCTAT |
| Core-178 | TTACCTTATGCGAAACCGGAAAAGACTTTTGCGGGTCTGGACGCGAGC |
| Core-179 | CACCCTTTTCAGCGGAGCGGAAAGCCTACCAAAATCACCA |
| Core-180 | GTTGCGCACATTTCAGTAGATCGCGGGGCCAGCTGGAGTAACA |

**Table S3 | Deviating DNA sequences of the core staples to enable the AuNP binding.** Staples Core-1, -59, -76, -94, -151, -153, -168, -171 and -175 were excluded and replaced by staples Core-181 to Core-185 and the AuNP-capture strands (Table S3).

| | |
|---|---|
| Core-181 | GAAGCCAGAGGAGAGGCAAGTTGG |
| Core-182 | GGGCCTCTTCGCTATCAGGGTTTAAAACCACGGAACTCAGGACGTTGGG |
| Core-183 | GGACCAGACGACTGGACCCGCTTGCCAGGGGTCTGGAGGCAAGGATAAAAATT |
| Core-184 | GAGAATCGGTTTGCGTCAATTCTACCTTTAAGAATAAAGT |
| Core-185 | ATAATTCGCGTCTGGGGATTCTCGGGAAACCTAGTAAACAAAGCGGCTGATGAAA |

**Table S4 | DNA sequences of the AuNP captur3 strands.** The AuNP-cap-3 strand additionally contains a 3-base pair overhang for monomer binding (green). The AuNP capture sequence is shown in red and the toehold region for the strand displacement reactions in blue.

| | |
|---|---|
| AuNP-cap-1 | CTGAATGGAAGAAGAAGAAGAGACACTAACGAGGACAGGCGAAACACACTAAAACAC TCAGGATTACCG |
| AuNP-cap-2 | CTGAATGGAAGAAGAAGAAGAGACACTAACTAGGCTGAACCAGAGAGAGTGCGAACGGCA AATGTGTACCA |
| AuNP-cap-3 | CTGAATGGAAGAAGAAGAAGAGACACTAACTACCAAGCATGAACAGGCGCAAGAAAAATC TAAC |
| AuNP-cap-4 | CTGAATGGAAGAAGAAGAAGAGACACTAACACAACGGACCAACTTCCTTCATATACCAGA ACATTATTACA |

**Table S5 | DNA sequences of connector strands I.** The 3-base pair overhangs are shown in green.

| | |
|---|---|
| Connector-I-1 | CAAGGCGATTTTTTGCAAACCA |
| Connector-I-2 | AATGTGAGCCATTAATGAATCG |
| Connector-I-3 | TAATACTTTTGGTAATCGTAAAACTTTA |
| Connector-I-4 | GCCAACGTTAGTTTGAAGA |
| Connector-I-5 | ATGTCAATCATGTAGCCAGCTTTCATCAACAAGC |
| Connector-I-6 | AGTTTTGCAAACTCCAA |
| Connector-I-7 | GCTGGCGAAAGGGGGATGTGCTG |
| Connector-I-8 | CAGGTCATCTTTGACCCCAG |
| Connector-I-9 | AAAACATTATGACCCTG |
| Connector-I-10 | TTAGATCGACAATGACACG |

**Table S6 | DNA sequences of connector strands IIa.** The 9-base pair overhangs are shown in blue, the blocking sequences are highlighted in red.

| | |
|---|---|
| Connector-IIA-1 | AAACGTAGATTCATTATAGCGTTTG |
| Connector-IIA-2 | AAGGTGAATTATCACGGAAATTAAAATACATACATA |
| Connector-IIA-3 | TTATCATTTTGCGGAACAAAGAAAAATAAAGAAATATACACAGTACATAA |
| Connector-IIA-4 | GTAGGAATCATTACCGCGCCCAGAGCCTAATTTGCCAGTTACAAAA |
| Connector-IIA-5 | ATTCTGTCCAGACGACGACAATAAGCCGTTTTTATTTTCATCATATCAGAG |
| Connector-IIA-6 | ACCTTGCTTCTGTAAATCGTCGCTATATGTAAATGCTGATGCAAATCCAAT |
| Connector-IIA-7 | GTTACAAAATAAACAGCCTATTGAC |
| Connector-IIA-8 | AAATCCAATCGCAAGACAAAAGAGGGTAATTGAGCGCTAGTAGGAATC |
| Connector-IIA-9 | ATATCAGAGAGATAACCCACTTATTACGCAGTATGTTAGCAGGTCAGAC |
| Connector-IIA-10 | ATCAATATATGTGAGTGAATAAGGTTTAAC |
| Connector-IIA-11 | TAGCGTTTGCCATCTTTCATAGCCCCCTTATAAGGTGAAT |
| Connector-IIA-12 | AGGTTTAACGTCAGATGAATTGCGTAGATTTTCACCTTGCTT |
| Connector-IIA-13 | AGTACCGACAAAAGGTAAAGTATTATCATTT |
| Connector-IIA-14 | TAAAAGTTTGAGTAACAATTCTGTCC |
| Connector-IIA-15 | AGGTCAGACGATTGGGACAGGAGGTTGAGGC |

**Table S7 | DNA sequences of connector strands IIb.**

| | |
|---|---|
| Connector-IIB-1 | TCAATATATGTGAGTGAATA |
| Connector-IIB-2 | AGATAACCCACTTATTACGCAGTATGTTAGC |
| Connector-IIB-3 | CCATCTTTCATAGCCCCCTTAT |
| Connector-IIB-4 | TGCGGAACAAAGAAAAATAAAGAAATATACACAGTACATAAA |
| Connector-IIB-5 | TATCACGGAAATTAAAATACATACATA |
| Connector-IIB-6 | CGCAAGACAAAAGAGGGTAATTGAGCGCTA |
| Connector-IIB-7 | GTCAGATGAATTGCGTAGATTTTC |
| Connector-IIB-8 | AGACGACGACAATAAGCCGTTTTTATTTTCATC |
| Connector-IIB-9 | AGTACCGACAAAAGGTAAAGTA |
| Connector-IIB-10 | CTGTAAATCGTCGCTATATGTAAATGCTGATGC |
| Connector-IIB-11 | TAAACAGCCTATTGAC |
| Connector-IIB-12 | AAACGTAGATTCATTA |
| Connector-IIB-13 | ATTACCGCGCCCAGAGCCTAATTTGCCA |
| Connector-IIB-14 | TAAAAGTTTGAGTAACA |
| Connector-IIB-15 | GATTGGGACAGGAGGTTGAGGC |

**Table S8 | DNA sequences of connector strands III.** The 3-base pair overhangs are shown in orange.

| | |
|---|---|
| Connector-III-1 | ACCAGGCGCAAGAAAAATCTAAC |
| Connector-III-2 | ATGTCAATCATGTAGCCAGCTTTCATCAACAAGC |
| Connector-III-3 | CAAGGCGATTTTTTGCAAACCA |
| Connector-III-4 | TAATACTTTTGGTAATCGTAAAACTTTA |
| Connector-III-5 | AAAACATTATGACCCTG |
| Connector-III-6 | TTAGATCGACAATGACACG |
| Connector-III-7 | AGTTTTGCAAACTCCAA |
| Connector-III-8 | CAGGTCATCTTTGACCCCAG |
| Connector-III-9 | GCCAACGTTAGTTTGAAGA |
| Connector-III-10 | GCGATTATACCAAGCATGAACGGTGTACCA |
| Connector-III-11 | GCTGGCGAAAGGGGGATGTGCTG |
| Connector-III-12 | AATGTGAGCCATTAATGAATCG |